\documentclass[a4paper,11pt]{article}
\pdfoutput=1 

\usepackage{jcappub} 

\usepackage[T1]{fontenc} 
\usepackage{diagbox}
\usepackage{comment}
\usepackage[official]{eurosym}

\title{\boldmath Needlet thresholding methods in component separation}

\author[a,b,1]{F. Oppizzi,\note{Corresponding author.}}
\author[b]{A. Renzi,}
\author[a,b,c]{M. Liguori,}
\author[d]{F. K. Hansen}
\author[e]{D. Marinucci}
\author[f,g]{C. Baccigalupi}
\author[a,b]{D. Bertacca}
\author[f]{D. Poletti}

\affiliation[a]{Dipartimento di Fisica e Astronomia "G. Galilei", Universit{\`a} degli Studi di Padova,\\ Via Marzolo 8, 35131 Padova, Italy}
\affiliation[b]{INFN, Sezione di Padova, via Marzolo 8, I-35131, Padova, Italy}
\affiliation[d]{Institute of Theoretical Astrophysics, University of Oslo, Blindern, Oslo, Norway}
\affiliation[e]{Dipartimento di Matematica, Universit{\`a} di Roma Tor Vergata}
\affiliation[f]{SISSA - Scuola Internazionale Superiore di Studi Avanzati,Via Bonomea 265, 34136, Trieste, Italy}
\affiliation[g]{INFN, Sezione di Trieste, Padriciano, 99, 34149 Trieste, Italy}

\emailAdd{filippo.oppizzi@pd.infn.it}
\emailAdd{alessandro.renzi@pd.infn.it}
\emailAdd{michele.liguori@pd.infn.it}
\emailAdd{frodekh@astro.uio.no}
\emailAdd{marinucc@mat.uniroma2.it}
\emailAdd{carlo.baccigalupi@sissa.it}
\emailAdd{daniele.bertacca@pd.infn.it}
\emailAdd{davide.poletti@sissa.it}

\abstract{Foreground components in the Cosmic Microwave Background (CMB) are sparse in a needlet representation, due to their specific morphological features (anisotropy, non-Gaussianity). 
This leads to the possibility of applying needlet thresholding procedures as a component separation 
tool.
In this work, we develop algorithms based on different needlet-thresholding schemes and use them as extensions of existing, well-known component separation techniques, namely ILC and template-fitting. We test soft- and hard-thresholding schemes, using different procedures to set the optimal threshold level. 
We find that thresholding can be useful as a denoising tool for internal templates in experiments with few frequency channels, in conditions of low signal-to-noise. We also compare our method with other denoising techniques, showing that thresholding achieves the best performance in terms of reconstruction accuracy and data compression while preserving the map resolution.
The best results in our tests are in particular obtained when considering template-fitting in an LSPE like experiment, especially for B-mode spectra.}

\begin{document}
\maketitle
\flushbottom
\section{Introduction}
\label{sec:intro}

The search for primordial polarization B-modes is the main and most exciting challenge for both the current and coming generation of Cosmic Microwave Background (CMB) experiments. 
A detection of B-mode CMB polarization would carry huge implications for our understanding of the Early Universe, essentially allowing for a smoking-gun confirmation of Inflation, as well as for the measurement of its energy scale \citep{Kamionkowski:1996ks,Zaldarriaga:1996xe,Seljak:1996gy,Kamionkowski:1996zd}.
At the same time, the quest for B-modes presents formidable challenges. The polarization signal from the inflationary stochastic Gravitational Wave (GW) background is expected to be very faint. 
Its detection, if achievable, will require an exquisite and unprecedented level of accuracy in controlling systematic biases in the data. One of the major sources of systematic contamination, besides instrumental effects, is the astrophysical foreground \citep{2018arXiv180104945P,doi:10.1063/1.3160888,PhysRevLett.114.101301,2016JCAP...03..052E,2018JCAP...04..023R,2018ApJ...853..127H,2016PhRvD..94h3526S,Delabrouille2009}.\\
A variety of methods have been so far designed, with the purpose to separate the CMB signal from foreground components. 
Of course, state-of-the-art component separation methods work very efficiently on the best available datasets, such as {\it Planck} \citep{2018arXiv180706208P,2018arXiv180104945P}. 
The target sensitivity for future B-mode surveys is, however, orders of magnitude below that of current experiments \citep{2019JLTP..194..443H,2019arXiv190704473A}. This -- together with the morphological complexity and incomplete understanding of polarized foregrounds -- makes further study and advancements in this area still crucial.

Different component separation algorithms exploit characteristic features of foreground emission to disentangle them from the background radiation, (mostly, but not only, their non-blackbody spectrum) \citep{2008arXiv0803.1814C,2012MNRAS.419.1163B,2004ApJS..155..227E,0004-637X-676-1-10,Bobin:2007hf,2012MNRAS.420.2162F,Baccigalupi:2000xy}.\\
In this work, we present an investigation on a technique relying on the assumption that the foreground signal is ``sparse'' in a proper representation, {\em i.e.} the majority of the signal is concentrated in few expansion coefficients.\\
To this purpose, we will rely on a needlet expansion of the CMB map. Needlets are a special kind of spherical wavelets, directly defined in harmonic space and not relying on any tangent plane approximation (see next section for details). They were developed as a functional analysis tool by \citep{Narcowich2006LocalizedTF} and applied to CMB analysis in various works \citep{baldi2009,baldi2009B,2006PhRvD..74d3524P,2013MNRAS.435...18B,2012MNRAS.419.1163B}. 
Like all wavelets, needlets display the property to be localized both in space (or time) and frequency, this is a key property to induce sparsity in the representation of coherent signals, such as foreground emission. In our analysis, we will exploit sparsity to either reconstruct foreground templates via thresholding procedures or for template denoising purposes. The idea is that these approaches can be combined with -- and be used as preliminary steps in -- different standard component separation methods: for example, initial template denoising in needlet space can be followed by standard template fitting; alternatively, needlet thresholding can be used for pre-cleaning single frequency channels, by exploiting morphological information (non-Gaussianity and anisotropy of foregrounds); the channels can then be combined with a common approach such as Internal Linear Combination (ILC). 
The idea is therefore not that of developing a component separation method per se, but rather to explore a range of applications in a general context. 
We will specify case by case what are the main novelties introduced in the different techniques which we are going to explore, and in which regime they find their best application. Our general finding is that the methods we explore here can be useful in situations characterized by limited frequency coverage, as it can be the case for current and forthcoming ground-based and balloon experiments. 
The paper is structured as follows: in section \ref{sec:need} we will review the main characteristic of the spherical needlet and we will introduce the notions of sparsity and thresholding. In section \ref{sec:method} we will show how these properties can be exploited for component separation and we will describe the techniques developed in this work. In section \ref{sec:resu} we will check the performance of our thresholding methods on simulations of different CMB surveys, showing a comparison with alternative techniques for template denoising, as well as possible applications in which thresholding is used in synergy with other methods. Finally, section \ref{sec:conc} is dedicated to the conclusions.

\section{Needlet Regression}
\label{sec:need}
Spherical wavelets are usually constructed by relying on a local flat sky approximation. 
This means that the basis function is defined on a flat tangent plane and then implemented on the sphere. 
The needlet basis, instead, is defined directly in harmonic space, in terms of spherical harmonics. 
As we will show in the following, this is a great advantage for the exact computation of the needlet coefficients.

More specifically, needlets are defined starting from the window function $b(\ell,j)$ that sets the harmonic support for each needlet layer $j$. 
This function must satisfy three properties: 
\begin{enumerate}
    \item compact support: $b(\ell,j)>0 \ {\rm if} \ \ell_{min,j}\leq \ell \leq \ell_{max,j} $ and $b(\ell,j)=0$ otherwise. This ensure that each needlet layers represents a fixed range of scales. Moreover, each layer $j$ will have equal support in $\log(\ell)$.
    \item partition to unity, so that $\sum_j b^2(\ell,j)=1$.
    \item smoothness, {\it i.e.} $b(\ell,j)$ is infinitely differentiable.
\end{enumerate}

Given a window function with these properties (see \citep{marinucci_peccati_2011} for a complete derivation), the spherical needlet basis function can be defined as:
\begin{equation}
    \psi_{jk}(x)=\sqrt{\lambda_{jk}}\sum_{\ell}b(\ell B^{-j})\sum_{m=-\ell}^\ell Y_{\ell}^{m}(\xi_{jk})\overline{Y}_\ell^m(x),
    \label{eq:need}
\end{equation}
where $j$ represents the needlets scale, $\lambda_{jk}$ and $\xi_{jk}$ are respectively the weights and the cubature points at the level $j$, and B is a parameter fixing the $\ell$ coverage of the needlets layers. 
As shown in \citep{2008MNRAS.383..539M}, the support of $b(\ell B^{-j})$ extends over $\ell \in (B^{j-1},B^{j+1})$, this is the multipoles window spanned by the needlets layer $j$.
From Eq. \ref{eq:need} we can define the needlets coefficients $\beta_{jk}$ of a square integrable function on the sphere $f(x)$ as:
\begin{equation}
    \beta_{jk}=\int_{S^2}{\rm d}x \ f(x)\psi_{jk}(x)=\sqrt{\lambda_{jk}}\sum_{\ell}b(\ell B^{-j})\sum_{m=-\ell}^\ell a_{\ell m}Y_{\ell}^{m}(\xi_{jk}),
    \label{eq:needtr}
\end{equation}
where we use:
\begin{equation}
    a_{\ell m}=\int_{S^2}{\rm d}x \ f(x)\overline{Y}_\ell^m(x).
\end{equation}
Equation \eqref{eq:needtr} is the direct needlet transform, and $\beta_{jk}$ are the needlet coefficients at scale $j$ in the position defined by the cubature point $\xi_{jk}$. 
The harmonic coefficients $a_{\ell m}$ can be computed numerically with a high level of precision \citep{1999astro.ph..5275G}, this makes the numerical implementation of needlets very convenient.\\
The inverse of equation \eqref{eq:needtr} is:
\begin{equation}
    f(x)=\sum_{jk}\beta_{jk}\psi_{jk}(x).
    \label{eq:invtr}
\end{equation}

With these preliminary definitions in hand, let us now review the main properties which make needlets a particularly suitable choice for CMB analysis. 
Needlets owe their name to their localization properties. The basis functions are localized quasi-exponentially around their centers, represented by the cubature points $\xi_{jk}$.
In their seminal paper \citep{Narcowich2006LocalizedTF}, Narcowich and collaborators proved this statement showing that for any point $x$ on the sphere surface, there exists a constant $c_{M}$, such that:
\begin{equation}
    |\psi_{jk}|\leq\frac{c_{M}B^j}{(1+B^{j}\arccos{(\xi_{jk},x}))^M}.
\end{equation}
Note that the function $\arccos$ in the above formula represents the distance on the sphere; this states exactly that the function $\psi_{jk}$ decreases faster than any power law. 
This property is of significant importance in CMB analysis, where the presence of missing observations poses a problem for the computation of harmonic coefficients due to the onset of spurious correlations between harmonic coefficients $ a _ {\ell m} $. 
These correlations represent a limitation for the evaluation of the power spectrum and the other cumulants and must be corrected for, generally with high computational costs. 
As proven in \citep{2008MNRAS.383..539M}, needlet coefficients are instead much less sensitive to gaps in the map, therefore working in needlet space allows avoiding to correct for missing observation.\\
Needlets are also particularly well suited for the representation of random fields on the sphere, due to their uncorrelation properties.
The fact that the window function $b(\ell B^{-j})$ has compact support in $(B^{j-1},B^{j+1})$ indeed ensures that theoretical correlations between $\beta_{jk}$ cancel if the difference in levels is greater than 2, so that if $j-j'>2$ we have:
\begin{equation}
    \beta_{jk}\beta_{j'k'}=\sqrt{\lambda_{jk}\lambda_{j'k'}}\sum_{\ell\ell'}b(\ell B^{-j})b(\ell' B^{-j'})\sum_{m m'}a_{jk}a_{j'k'} Y_{\ell}^{m}(\xi_{jk})Y{\ell'}^{m'}(\xi_{jk})=0,
\end{equation}
as the simple consequence of the fact that the supports of the two basis functions do not overlap.
If we consider instead fixed scale correlations, it was proven in \citep{baldi2009} that the needlet representation of a Gaussian random field with smooth power spectrum satisfies:
\begin{equation}
    |Corr(\beta_{jk}\beta_{jk'})|=\left|\frac{\beta_{jk}\beta_{jk'}}{\sqrt{\beta_{jk}^2\beta_{jk'}^2}}\right|\leq\frac{c_M}{(1+B^{j}\arccos{(\xi_{jk},\xi_{jk'}}))},
\end{equation}
for any positive integer M and $c_M>0$. 
This implies that, for growing scale $j$, needlets coefficients are asymptotically uncorrelated.
Therefore, under Gaussianity, the small scales coefficients behave approximately as a sample of i.i.d. random variables. 

\subsection{Sparsity and Thresholding}
\label{sec:stlap}
A signal is said to be sparse if -- in a given basis or frame -- it can be reconstructed by using only a small amount of basis elements (see {\it e.g.} \citep{Starck:2015:SIS:2901581} for a complete review on the subject).
Wavelets and needlets present several crucial features which allow for sparse representation of signals.
First, they are not an orthonormal basis but instead a tight frame. This implies that the basis contains redundant elements.
Furthermore, their tight space-frequency localization properties make them suitable to identify discontinuities in the signal and represent them with just a small number of modes. 
In such conditions, sparsity can always be achieved if the signal under study is smooth (although this is not a necessary condition). 
This makes sparsity a key element in wavelet regression methods \citep{doi:10.1093/biomet/81.3.425}, since it allows developing efficient techniques to separate a coherent signal (which is sparse, for the reasons just mentioned above) from a stochastic ``noise'' component.
Note that in our setting, CMB itself can be viewed as part of this ``noise'' (for the purpose of foreground estimation).
In general, stochastic fields do not admit a sparse representation, due to their lack of smoothness and, in the case of CMB, to its isotropy.
Such separation can be essentially achieved by setting to zero all the coefficients under a certain threshold and eventually rescaling the remaining ones. 
This clearly filters the few large signal coefficients from the noise background. 
Such a procedure is referred to as thresholding and it is, in spirit, similar to principal component analysis, since it aims to reduce the complexity of a multidimensional data-set, by identifying the most significant modes.\par
The simplest thresholding scheme is called hard thresholding (HT); the effects of the hard thresholding operator on the needlet coefficients are simply:
\begin{equation}
    HT(\beta_{jk})=\begin{cases}0 &\quad {\rm if} \quad |\beta_{jk}|<\lambda\\
                                \beta_{jk} &\quad {\rm if} \quad |\beta_{jk}|\geq\lambda ,
    \end{cases}
\end{equation}
where $\lambda$ is a given threshold. 
In the case of a coherent signal, only few significant coefficients survive this operation, providing an optimal representation as well as efficient data compression.\\
A second option is soft thresholding (ST). In this case, the significant coefficients are rescaled, proportionally to the chosen threshold:
\begin{equation}
    ST(\beta_{jk})={\rm sgn}(\beta_{jk})(|\beta_{jk}|-\lambda)_{+}\begin{cases}\beta_{jk}+\lambda &\quad {\rm if} \quad \beta_{jk}\leq\lambda\\
    0 &\quad {\rm if} \quad |\beta_{jk}|<\lambda\\
                                \beta_{jk}-\lambda &\quad {\rm if} \quad \beta_{jk}\geq\lambda ,
    \end{cases}
\end{equation}
where the operator $(*)_+$ stands for the positive part of the argument.\\
It is known that the soft thresholding solution can be interpreted -- from a Bayesian perspective -- as a maximum posterior estimator from a Gaussian Likelihood with a leptokurtic Laplace prior on the parameters, which in our case are the needlet coefficients. 
Let us briefly show this.
Consider a dataset $x=\theta+n$, where $\theta$ is a signal that is sparse in some basis and $n$ is a Gaussian noise with known variance $\sigma^2$.
Assume also that the scale parameter $1/\lambda $ of the Laplace prior on $\theta$ is known and its mean is 0, so that we can write:
\begin{align}
    P(\theta|x)&\propto\mathcal{L}(x|\theta)P(\theta)=N(x;\theta,\sigma)L(\theta;\lambda,0), \\
    -\log{P(\theta|x)}&= \frac{(x-\theta)^2}{2\sigma^2}+\lambda|\theta|+const, \label{eq:post}
\end{align}
where we use $N(*;\mu,\sigma)$ and $L(*;\mu,\lambda)$ to define respectively the Normal and the Laplace distributions, that is:
\begin{equation}
    L(d;\lambda,\mu)=\frac{\lambda}{2}e^{-\lambda|d-\mu|} .
\end{equation}
The maximum posterior estimator (MPE) is obtained by minimizing equation \eqref{eq:post}.
We start by taking the derivative with respect $\theta$ (that we denote with $\partial_{\theta}$):
\begin{align}
    \partial_{\theta}(-\log{P(\theta|x)})&=-\frac{(x-\theta)}{\sigma^2}+\lambda\partial_{\theta}|\theta|=0,\\
    \theta&=x-\sigma^2\lambda\partial_{\theta}|\theta| \label{eq:mpelap},
\end{align}
since the absolute value is not differentiable around zero (and equivalently the L1 norm $||\theta||$, if  dealing with multidimensional data), we should take the subgradient, so that we have:
\begin{align}
    \partial_{\theta}||\theta||=\begin{cases}
    1 & \quad \text{if } \theta>0\\
    -1 & \quad \text{if } \theta<0\\
    [-1,1] & \quad \text{if } \theta=0,
  \end{cases}
  \label{eq:subl1}
\end{align}
note that for $\theta=0$ the subgradient is actually an interval of values. 
We can understand the soft thresholding solution applying the conditions \eqref{eq:subl1} at equation \eqref{eq:mpelap}. 
First notice that the term $\sigma^2\lambda\partial_{\theta}|\theta|$ can only take values in the interval $[-\sigma^2\lambda,\sigma^2\lambda]$.
Therefore,considering the case $|x|> \sigma^2\lambda$ we must have $\theta\neq0$ to satisfy condition \eqref{eq:mpelap}.
From the same equation we see that it must be ${\rm sgn}(\theta)={\rm sgn}(x)$. 
In this case, the solution is given by  $\theta=x-{\rm sgn}(x){\sigma^2\lambda}$.
For $|x| = \lambda \sigma^2$ instead, $P(\theta|x)$ is maximum in $\theta=0$ since ${\rm lim}_{x\to\sigma^2\lambda}(x-{\rm sgn}(x){\sigma^2\lambda})=0$ (remind that $P(\theta|x)$ is continuous).
At last, if we have $|x|< \sigma^2\lambda$ , the only admissible solution of \eqref{eq:mpelap} is $\theta=0$ otherwise, due to condition \eqref{eq:subl1}, it would give $\theta<0$ for $\theta>0$ and vice versa.
After these considerations it is clear that the solution coincides with the soft thresholding operator, that in this case is:
\begin{equation}
    ST(x)={\rm sgn}(x)(|x|-\sigma^2\lambda)_{+}.
\end{equation}

\section{Methodology}
\label{sec:method}
\begin{figure}
    \centering
    \includegraphics[width=1.\textwidth]{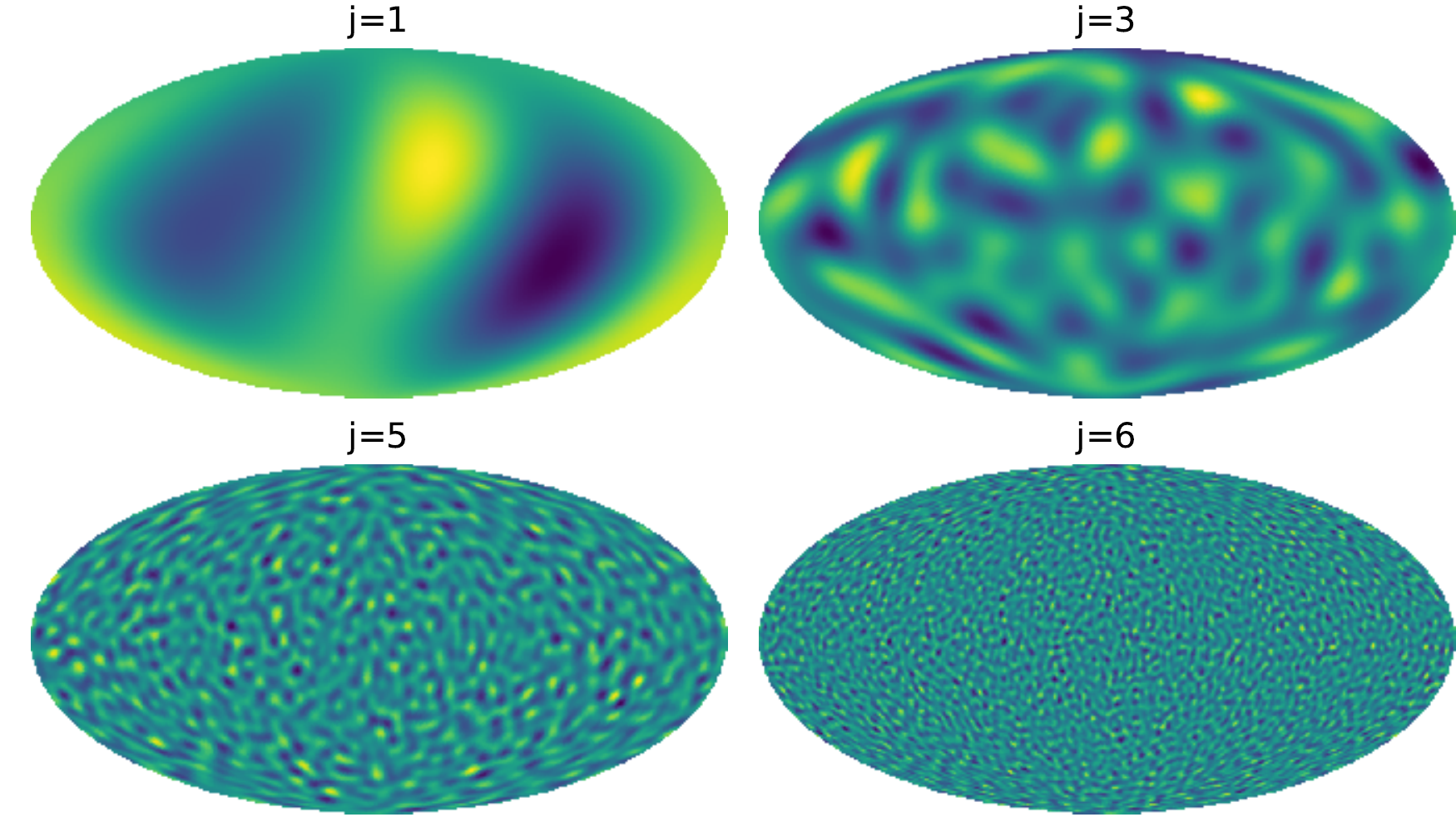}
    \caption{Needlet coefficients of a Gaussian random realization for some needlet scale j with  scale parameter B=2, the corresponding multipole coverage is: $\ell=[1,4]$ for $j=1$, $\ell=[4,16]$ for $j=3$, $\ell=[16,64]$ for $j=5$, $\ell=[32,128]$ for $j=6$.}
    \label{fig:cmbneed}
\end{figure}
The general idea behind this investigation is that foreground signals and the CMB fluctuations can be disentangled when the data are represented in a proper basis, frame or dictionary, and that needlets represent an ideal choice to this purpose.
Foreground emission comes, in the larger part, from coherent sources concentrated around the galactic plane and in few large structures that extend at higher galactic latitudes. 
As we saw in the previous sections, a space-frequency representation of coherent signals naturally tends to be sparse. 
We will thus expect that the contribution from the galactic foreground will be concentrated in few large coefficients that can be identified and fitted with a needlet thresholding technique.
On the other hand, CMB has very different features, since it is a random isotropic field and not a coherent signal.
Thus, unlike foregrounds, a needlet-space representation of the CMB signal will not be sparse. 
The reason is that the CMB mostly does not form coherent structures, but it is instead a homogeneous fluctuations field at all scales.\par
\begin{figure}
    \centering
    \includegraphics[width=1.\textwidth]{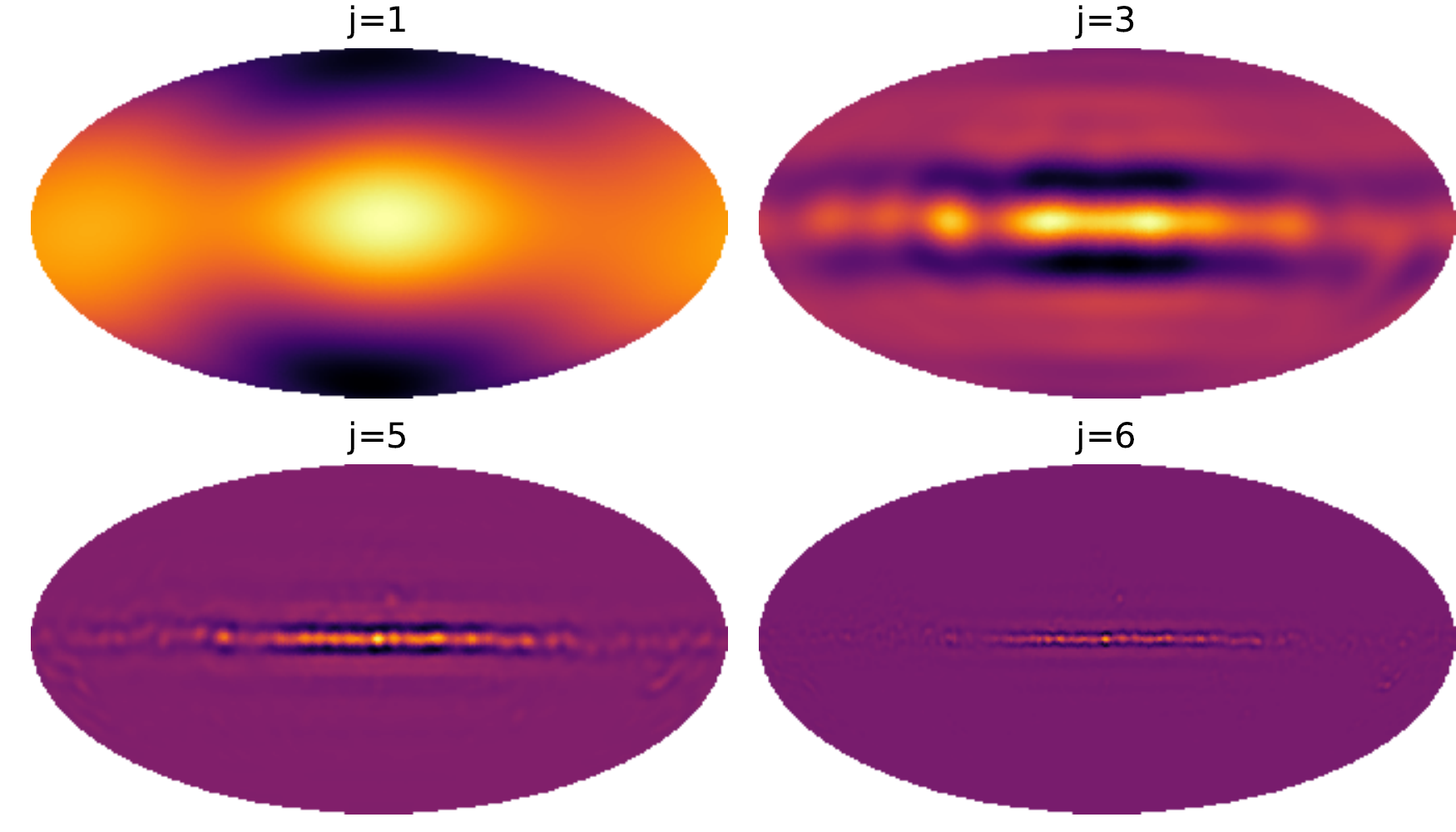}
    \caption{Needlet coefficients of thermal dust template for some needlet scale j with  scale parameter B=2, the corresponding multipole coverage is: $\ell=[1,4]$ for $j=1$, $\ell=[4,16]$ for $j=3$, $\ell=[16,64]$ for $j=5$, $\ell=[32,128]$ for $j=6$.}
    \label{fig:dustneed}
\end{figure}

The different behaviour of the two components can be immediately appreciated by looking at the needlet decomposition of a CMB realization and a foreground template. 
We show in figure \ref{fig:cmbneed} the needlet decomposition of a random CMB realization: the signal is spread over all the coefficients, as expected, given its stochastic nature.
Furthermore, it is easy to notice that only adjacent layers show some level of correlation.
Needlets split a continuous field in several independent realizations, the layers, each one covering a limited range of frequencies.
Figure \ref{fig:dustneed} shows instead the corresponding decomposition of a thermal dust template.
We see how the information is actually concentrated only in few coefficients located near the galactic plane.
The lowest layers trace the diffuse emission while the higher frequency levels contain only few small scale corrections.
Moreover, all scales are highly correlated at any distance in frequency level, and this is expected since the signal is coherent.\par
To get further insight into how the thresholding procedure operates in a component separation context, we now illustrate and justify it within a general Bayesian framework.
As shown in e.g., \citep{2016A&A...588A.113V},  a Bayesian approach provides a way to describe different component separation techniques within a unified, general formalism. This approach shows in particular how different common component separation methods amount to different choices of priors and marginalized parameters.\par
We start as usual by assuming to have observations from $K$ channels, with a mixture of $N$ components and $M$ elements (pixels) in each channel (by ``pixel'' we mean real space pixels, $a_{\ell m}$, needlet coefficients or the elements of whichever basis is adopted to represent the signal).  We recall then the linear mixture model:
\begin{equation}
    d_i=As_i+n_i,
    \label{eq:linmix}
\end{equation}
where $d_i$ is a vector of $K$ elements representing the observations in a given pixel $i$, $s_i$ is a vector of $N$ elements representing the contribution of the components in the same pixel, $A$ is the {\it mixing matrix} of dimension $N\times K$, which weighs the contributions of different components at different frequencies; finally, $n_i$ is the noise in the pixel $i$.\par
The Bayesian formulation of the component separation problem aims to solve:
\begin{equation}
    P(A,s|d)\propto\mathcal{L}(d|A,s)P(s),
\end{equation}
as shown e.g., in \citep{2016A&A...588A.113V}. 
With specific assumptions on the priors and eventually variables to marginalize over, the formulation above can be used to define typically adopted component separation techniques, such as ILC and SMICA \citep{2008arXiv0803.1814C,2005EJASP2005..100M}.
In our case, we want to introduce the sparsity assumption on the foreground templates. 
A similar hypothesis is also at the basis of the development of other algorithms as, for example, GMCA \citep{Bobin:2007hf}. We will soon clarify the differences between the approach discussed here and GMCA.

As noticed in the previous section, the usual way to enforce sparsity in a Bayesian context is to assume leptokurtic priors, with the Laplace distribution as a common choice.
We thus want to study the implementation of this kind of prior in a Bayesian component separation framework.\par
First, we must rewrite the linear mixture model in a form more suited to our scope, which is ultimately that of recovering the CMB signal.
We assume that the foreground signal is sparse in the needlet domain, but the CMB is not, this is the main difference with the GMCA algorithm, which instead implicitly assumes sparsity also for the CMB component.
In more detail, GMCA aims at the reconstruction of the full mixing matrix $A$ and the signal $s$ by exploiting the morphological diversity of the components in a given basis \citep{refId0,fadili:hal-00808047}.
In this work, instead, our purpose is to isolate the cosmological stochastic signal from the other emissions, avoiding the full reconstruction of the components.\\
We thus rewrite the linear mixture model assuming that components with a ``Gaussian'' and a ``non-Gaussian'' prior probability coexist in the data. Besides noise, we assume that the only ``Gaussian'' component is the CMB; thus we have, in the single pixel:
\begin{equation}
    d_{i}= As_{i}+{\bf e} c_i + n_i=f_i+{\bf e}c_i+n_i,
    \label{eq:linmix2}
\end{equation}
where we explicitly separate the CMB from the other components, by denoting it with $c_i$, times the vector of ones ${\bf e}$ of length $K$ (since the CMB signal is constant between channels). 
Furthermore $f_i=A's_i$, where $A'$ is the mixing matrix with the row corresponding to the CMB set to zero.
We rewrite the model in this way because at this stage we are not interested in the mixing matrix $A$. Therefore instead of explicitly estimating the underlying $N$ templates, we consider the linear combination, $f_i$, of all of them, in each of the $K$ channels. 
We assume that a Laplace distribution is a proper prior also for this combination.\par
\noindent We now split the problem as follows:
\begin{align}
    &\begin{cases} P(f,c|d) = P(c|f,d)P(f|d),  \\
    P(f|d) \propto P(f)\int_{\infty} {\rm d}c \ \mathcal{L}(d|c,f)P(c), 
    \end{cases}   \label{eq:margpost} \\
    &\to P(f,c|d) \propto P(c|f,d)P(f)\int_{\infty} {\rm d}c \ \mathcal{L}(d|c,f)P(c). \label{eq:postpost}
\end{align}
We can then solve the problem of recovering the CMB component in two steps. We first find the value $\hat f$ which maximize \ref{eq:margpost}, followed by replacing such value in \ref{eq:postpost} and maximizing again.\\
Since the noise component is uncorrelated between channels and since in our approach we are not interested in recovering explicitly the full mixing matrix, we can assume we will repeat our thresholding procedure independently at each frequency. 
Therefore, in our present derivation, we treat $d,f$ and $c$ as single-channel maps. 
The data likelihood takes the usual form (Gaussian noise):
\begin{align}
    &\mathcal{L}(d|c,f)=N(d;c+f,C_n)\propto \exp\left[-\frac{1}{2}(d-f-c)^{\bf^T}C_n^{-1}(d-f- c) \right]\\
    &P(c)=N(c;0,C_c)\propto\exp\left[-\frac{1}{2}c^{\bf^T}C_c^{-1}c \right],
\end{align}
where $C_n$ and $C_c$ are respectively the noise and CMB covariance matrices. Therefore we can write:
\begin{align}
    \mathcal{L}(d|c,f)P(c)&\propto\exp\left[-\frac{1}{2}d^{\bf^T}C_n^{-1}d+f^{\bf^T}C_n^{-1}d-\frac{1}{2}f^{\bf^T}C_n^{-1}f\right] \times \nonumber\\
    & \times \exp\left[-\frac{1}{2}c^{\bf^T}\left(C_n^{-1}+C_c^{-1}\right)c+c^{\bf^T}\left(C_{n}^{-1}d-C_{n}^{-1}f \right)\right],
\end{align}
Marginalization over $c$ can be carried out via standard Gaussian integration. We remind:
\begin{align}
\int \exp\left[-\frac{1}{2}\vec{x}^T \bold{R} \vec{x}+\vec{B}^T \vec{x}\right] d^nx= \sqrt{ \frac{(2\pi)^n}{\det{R}} }\exp\left[\frac{1}{2}\vec{B}^{T}\bold{R}^{-1}\vec{B}\right],
\end{align}
to obtain:
\begin{align}
    \mathcal{L}(d|f)=&\int_{\infty} {\rm d}c \ \mathcal{L}(d|f,c)P(c)\propto\exp\left[-\frac{1}{2}d^{\bf^T}C_n^{-1}d+f^{\bf^T}C_n^{-1}d-\frac{1}{2}f^{\bf^T}C_n^{-1}f\right]\times \nonumber \\
    &\times\exp\left[\frac{1}{2}\left(C_{n}^{-1}d-C_{n}^{-1}f \right)^{\bf^T}\left(C_n^{-1}+C_c^{-1}\right)^{-1}\left(C_{n}^{-1}d-C_{n}^{-1}f \right) \right].
    \label{eq:marlik}
\end{align}
We then implement the Laplacian prior, again assuming uncorrelation between the channels. Thus we have $P(f)\propto \exp{-\lambda||f||}$ where $||*||$ is the L1 norm. 
We add this to (\ref{eq:marlik}) define $R \equiv \left(C_n^{-1}+C_c^{-1}\right)$ for simplicity of notation. Finally, we differentiate posterior with respect to the foreground component, to find:
\begin{equation}
    \partial_f(-\log(P(f,d)))=-C_{n}^{-1}d+C_{n}^{-1}f+C_{n}^{-1}R^{-1}C_{n}^{-1}d-C_{n}^{-1}R^{-1}C_{n}^{-1}f+\lambda\partial_f{||f||},
\end{equation}
which implies:
\begin{equation}
   \hat f=d-\left(C_{n}^{-1}- C_{n}^{-1}R^{-1}C_{n}^{-1}\right)^{-1}\lambda\partial_f{||f||}.
\end{equation}
As derived in section \ref{sec:stlap}, the solution to this problem is the soft thresholding operator, with threshold $\left(C_{n}^{-1}-C_{n}^{-1}R^{-1}C_{n}^{-1}\right)^{-1}\lambda$.
Following the above derivation, the central idea of our study is therefore that of using thresholding as a preliminary tool for foreground cleaning or foreground template reconstruction, in single channels. 
This captures morphological information in the foreground spatial distribution, at any fixed frequency. 
Channels can then be combined and the overall cleaning procedure further refined by applying standard algorithms that exploit CMB and foreground spectral properties. 
To this purpose, in the following, we combine thresholding with template fitting and Internal Linear Combination.
Before concluding this section, let us stress here that our thresholding operator is applied to reconstruct foregrounds and not the CMB component. Reconstructing the CMB via thresholding in specific representations could break isotropy and Gaussianity, and we explicitly avoid this potential issue with our approach \citep{Cammarota:2013wba,Feeney:2013bqa}.

\subsection{Template Fitting}
Template fitting provides estimates of the amplitudes of each component from the fit of known templates to the data of interest. 
The results are then subtracted from the data to remove the spurious signal.\par
In the linear mixture model \eqref{eq:linmix}, the distribution of the components over the data is stored in the matrix ${s}$.  
The templates used should then reproduce the elements of $ { s} $, other than the CMB, or a linear combination thereof.
Assuming to have the exact templates, the linear fit would provide the entries of the mixing matrix ${ A}$ corresponding to the given source and channel.
The result for a collection of $N_{\rm temp}$ known templates ${ T}$ fitted to a map ${ y}$, is the standard linear regression solution. 
Calling $\boldsymbol{\alpha}$ the vectors of estimated amplitudes we have:
\begin{equation}
    \boldsymbol{\alpha}=\left({ T}^{\rm T}{ C}^{-1}{ T} \right)^{-1} \left({T}^{\rm T}{ C}^{-1}{y} \right),
    \label{eq:tempfit}
\end{equation}
where ${C}$ is the $N_{\rm pix}\times N_{\rm pix}$ covariance matrix of the map, which depends on the noise and the CMB, and $T$ is the $N_{\rm pix}\times N_{\rm temp}$ matrix representing the template. 
The estimation and the inversion of ${ C}$ is a major limitation in template fitting since, given the high number of data points collected by modern surveys, it is very large and computationally expensive.\par
The choice of templates is, obviously, the crucial part of any template fitting technique.
A possibility is that of resorting to external templates from previous experiments or theoretical modeling. 
This approach requires a lot of a-priori information on the emissions of interest, which can be unavailable or unreliable.
Relying on external data-sets also runs into the issue of having to deal with additional systematic effects, cross-calibration problems and so on.
For this reason, most of the modern CMB surveys have been designed with several channels at foreground dominated frequencies, allowing us to track spurious contaminant components without having to rely on external information.
These templates obtained directly from the data are called internal templates. 
The most straightforward approach would be just to use these foreground dominated maps as templates, for example, a high-frequency channel as a dust template and a low frequency as synchrotron. 
However, these maps contain also the CMB contribution, which would be removed together with the contaminants, so that a correction factor must be introduced in formula \eqref{eq:tempfit}.
Another widely implemented solution is thus to build linear combinations of different channels so that the constant component (the CMB, providing that the observations are calibrated to its black body spectrum) vanishes   \citep{2012MNRAS.420.2162F,Hansen:2006rj}.
These combinations usually are computed as the subtraction of adjacent frequency channels. 
The major drawback of this approach is that the noise in the internal templates is enhanced compared with the single channel. 
{\em A natural application of the thresholding techniques just described is thus the denoising of the internal difference templates}.\par
In other words, our goal is to clean a ``target'' map with a given number of noisy foreground templates; a needlet thresholding algorithm is then used to obtain optimal templates in which the noise component is minimized while preserving the resolution of the starting template. 
In the next section, we will discuss in more detail the ``optimality'' of the thresholding solution for this purpose, and we will show a comparison with other viable estimators.
In our procedure, these templates are then fitted to the target channel, and the fitting coefficients are then used to combine the {\em original} templates.
The map obtained is an estimate of the foreground contribution in the target channel, which is then cleaned by subtracting it.
We start by decomposing the data (the target channel and the internal templates) in needlet coefficients.  
For each needlet scale $j$, the templates are then thresholded with threshold $\lambda_i$ and fitted to the data to obtain the amplitude coefficients $\alpha_i$ (where $i$ runs over the different templates). The optimal thresholds are selected in recursive way to maximize the goodness of fit of the templates with the target channel; namely, we iteratively threshold and fit the internal templates so that:
\begin{equation}
    \chi^2=\sum_k \frac{\left(\beta_{jk}^{map}-\sum_i\alpha_i\beta_{jk}^{T_i}(\lambda_i)\right)^2}{\sigma^2_{jk}},
\end{equation}
is minimized. Here, $\beta_{jk}^{map}$ represents the target channel at the needlet scale j (With k running over pixels), $\beta_{jk}^{T_i}(\lambda_i)$ are the templates thresholded with threshold $\lambda_i$ and the coefficients $\alpha_i$ are obtained with the standard template fitting solution. 
We will discuss other threshold selection method in the next section.\\ 
The templates are thus linearly combined with weights $\alpha_i$ and subtracted to the target map as a standard template fitting procedure.
We stress the fact that the cleaned template are used only to obtain the fitting coefficient, while the full templates are used to clean the map.
This allows us to preserve the linearity of the template fitting procedure. 
In section \ref{sec:tfres}, we will show the results of the application of this algorithm to different simulated data-sets.

\subsection{ILC}

As a further case study and an example of the versatility of our approach, we merge our thresholding algorithm with an Internal Linear Combination cleaning procedure. 
The general idea is as follows: a thresholding algorithm is ``inverted'' to remove from the map the most contaminated coefficients. This is followed by combining the channels, following the usual ILC prescription.
Since we work in needlet space, such method can be straightforwardly implemented in needlet-based ILC pipelines, such as NILC. However, this is by no means mandatory, and any other signal representation domain can be chosen in the ILC step.\par
The overall rationale of the approach is as usual that thresholding exploits  complementary foreground information, compared to ILC, since it allows us to minimize the foreground contribution in {\em single channels}, on the basis of spatial -- rather than spectral -- features. Merging the two methods could therefore in principle lead to useful improvements, especially in experiments where a limited number of frequency channels are available. 
In the section \ref{sec:ILCres} we will show the results of the application of our ILC with thresholding procedure, using simulated data-set and comparing it with the standard approach.\par
The algorithm developed in these tests is structured as follows.
First of all, each channel map is decomposed in needlet coefficients, and we treat different scales separately. In other words, we exploit the flexibility of the needlet representation to identify and remove large, spurious coefficients scale by scale. Note that, to avoid distortions in the spectral energy distribution of foregrounds, which would compromise the ILC step, in this analysis the masked coefficients after thresholding must be {\em the same in each channel}.
In practice, we construct an initial linear combination of the channels (for example with a preliminary ILC) to identify the foreground dominated regions as the isotropic residuals in this co-added map. We then go back to single-channel maps and remove these regions at each frequency. The criterion to choose the threshold is again recursive and based on the minimization of the anisotropy of the residual coefficients. More specifically, in the thresholding step, we minimize:
\begin{equation}
    \Delta_j=\frac{1}{N_{j}}\sum_k\left[\frac{\left(\beta_{jk}-\beta_{jk}^{T_{(\lambda)}}\right)^2}{\sigma^2_{j}}-1 \right]^2,
\end{equation}
where, as usual, $j$ set the scale, $\beta_{jk}$ represents the map in needlet space and $\beta_{jk}^{T_\lambda}$ is the map thresholded with threshold $\lambda$, $k$ is the position index (the HEALPix pixels, in our algorithm), $N_{j}$ is the number of coefficients at the given layer $j$ and the variance $\sigma_j$ can be estimated from the coefficients themself.
After this "pre-cleaning" via thresholding,  the different channels are then combined again with an ILC algorithm to produce a final CMB map.
The overall method can be essentially interpreted as a needlet space masking, where the masked area varies with the scale, followed by Internal Linear Combination.
At the end of this procedure, we extract the power spectrum from the cleaned maps, correcting for the missing coefficients with the standard MASTER technique \citep{Hivon:2001jp}.
The final power spectrum is used as a figure of merit to assess the performance of both our template fitting and ILC algorithms.
The performance of our algorithms is discussed in the next section.

\section{Results}
\label{sec:resu}
The initial goal of our analysis is that of verifying how well we can reconstruct noisy foreground templates at high resolution, using needlet thresholding. We then focus on the issue of combining our thresholding procedure with standard component separation methods - namely ILC and template fitting - with the aim of improving their performance. However, before moving to these points, we start by quantitatively checking the foreground sparsity assumption, on which the entire procedure is based.
\subsection{Measuring sparsity}

Our main goal is to separate the CMB and foreground components, by exploiting the sparsity of the latter. 
It is, therefore, useful to start our analysis by defining a metric to quantify sparsity and verify in detail whether and to which extent this assumption holds in our context of interest.\par
It has been proven that a good measure of sparsity is the so-called Gini index \citep{5238742,7071576}, defined as:
\begin{equation}
    G_P=1-2\int_0^1\mathrm{d}C(x) \ \frac{\int_0^x\mathrm{d}t \ t P(t) }{\int_0^\infty\mathrm{d}t \ t P(t)},
\end{equation}
where $P(x)$ is a positive valued probability distribution, with cumulative distribution $C(x)$.

For an ordered data set, $d=\{[d_1,...,d_i,...,d_N]$ : $d_i<d_{i+1} \forall \ i<N$\}, the Gini index can be estimated with the formula:
\begin{equation}
    \hat G(d)=1-2\sum_{i=1}^N\frac{d_i}{\sum_{k=1}^N d_k}\left(\frac{N-i+0.5}{N}\right).
\end{equation}
Originally, the Gini index was introduced in social economics to measure the degree of inequality of the income distribution of a population. It runs from $0$ (perfect equality, every person has the same wealth) to $1$ (perfect inequality, one person owns all the goods) \citep{AABERGE2001115}.
It is immediate to notice that the notion of inequality as defined above corresponds to the notion of sparsity in signal processing. To check the sparsity assumption, we thus measure the Gini index of different foreground templates and we compare it to a Gaussian realization.
Since the Gini index is defined for positive valued distributions, we use the square of the needlet coefficients $\beta_{jk}$.\par
\begin{table}
\centering
\begin{tabular}{|c||c|c|c|c|c|c|c|c|c|c|}
\hline $j$ & $2$ & $3$ & $4$ & $5$ & $6$ & $7$ & $8$ & $9$ & $10$ & $11$  \\
\hline $\ell_{peak}$ & $4$ & $8$ & $16$ & $32$ & $64$ & $128$ & $256$ & $512$ & $1024$ & $2048$  \\ 
\hline \hline {\bf CMB  }  & $0.669$ & $0.652$ & $0.633$ & $0.638$ & $0.637$ & $0.637$ & $0.637$ & $0.637$ & $0.637$ & $0.637 $ \\
\hline {\bf dust}          & $0.599$ & $0.788$ & $0.906$ & $0.957$ & $0.982$ & $0.991$ & $0.993$ & $0.995$ & $0.997$ & $0.997$  \\
$f_{sky}=0.8$        & $0.692$ & $0.693$ & $0.809$ & $0.886$ & $0.938$ & $0.969$ & $0.985$ & $0.993$ & $0.997$ & $0.997$ \\
\hline {\small {\bf syncrothron}}   & $0.688$ & $0.844$ & $0.933$ & $0.969$ & $0.986$ & $0.988$ & $0.979$ & $0.989$ & $0.994$ & $0.995$  \\
$f_{sky}=0.8$        & $0.719$ & $0.754$ & $0.812$ & $0.870$ & $0.895$ & $0.873$ & $0.844$ & $0.842$ & $0.849$ & $0.868$ \\
\hline {\bf AME} & $0.631$ & $0.823$ & $0.940 $ & $0.988$ & $0.998$ & $1.000   $ & $1.000$    & $1.000   $ & $1.000   $ & $1.000   $   \\
$f_{sky}=0.8$        & $0.687$ & $0.694$ & $0.830$ & $0.918$ & $0.954$ & $0.977$ & $0.995$ & $0.999$ & $0.999$ & $0.997$ \\
\hline {\bf free-free}     & $0.739$ & $0.868$ & $0.941$ & $0.974$ & $0.989$ & $0.995$ & $0.997$ & $0.997$ & $0.997$ & $0.998$ \\
 $f_{sky}=0.8$       & $0.711$ & $0.852$ & $0.935$ & $0.969$ & $0.983$ & $0.991$ & $0.995$ & $0.997$ & $0.997$ & $0.995$ \\
\hline
\end{tabular}
\caption{Gini index of the square of needlet coefficients $\beta_{jk}^2$ of various components computed scale by scale, both full-sky and outside the galactic plane ($f_{sky}=0.8$). Values approaching $1$ correspond to a more sparse representation. $0.637$ is the expected value for the $\chi^2$ distribution.} 
\label{tab:Gini}
\end{table}
Table \ref{tab:Gini} shows the results for intensity templates of different components, at different scales $j$, setting the needlet scaling parameter $B=2$ (we remind that each scale $j$ cover the multipole interval $[B^{j-1},B^{j+1}]$, and the corresponding window function peaks at $B^j$). We report the values computed both on the full sky and outside the galactic plane, with a galactic mask covering the $20\%$ of the sky. 
The first row refers to a random CMB realization. 
The following rows refer instead the principal sources of foreground emission. 
Since they roughly follow the shape of the Galaxy, the results are similar between components and are less sparse outside the galactic plane, where the diffuse emission dominates.
We generate the templates with the PySM software, based on the Planck sky model \citep{refId1,2017MNRAS.469.2821T}. \\
If we start by looking at the Gini index of the square of the CMB coefficients, we notice that it remains substantially constant between scales and it converges to $0.637$ for high $j$. 
This is the expected value for the $\chi^2$ distribution, {\it i.e.} the distribution of the squares of a Gaussian random variable. 
This behaviour reflects what stated in section \ref{sec:need}, namely that the needlet coefficients are asymptotically uncorrelated/independent at higher frequencies. 
If we now focus on the foreground components, the first thing we see is that lower frequencies display lower values of the Gini index. This is due to the fact that lower frequencies represent diffuse emission, which is not sparse.
Layers corresponding to higher multipoles are instead significantly more sparse, with a high degree of sparsity achieved at $j \geq 3$ or $\ell \gtrsim 10$ in the corresponding harmonic representation. This reflects the fact that, at smaller scales, we track single structures rather than a homogeneous fluctuations field. 
The small scale portion of the signal can thus be recovered keeping only the coefficients corresponding exactly to the size and positions of these structures, which are few compared to the total number of cubature points.\\
In light of these observations, we, therefore, expect our thresholding-based methods to achieve significantly better performance at $\ell \gtrsim 10$.\\
Let us point out that the fact that very large scales are not sparse does not refute in any way the overall sparsity assumption. 
This is of course due to the fact that the cardinality of the needlet coefficients at a given frequency scales as $B^{2j}$, thus the number of low $j$ $\beta_{jk}$ is negligible compared to the total.
Summarizing, we have just verified that foreground templates can be faithfully represented taking almost all the (few) large scale needlet coefficients and a minimal portion of the high-frequency ones, selected with a thresholding algorithm.
This highlights how thresholding can also achieve a remarkable level of data compression. 
\begin{figure}
    \centering
    \includegraphics[width=0.54\textwidth]{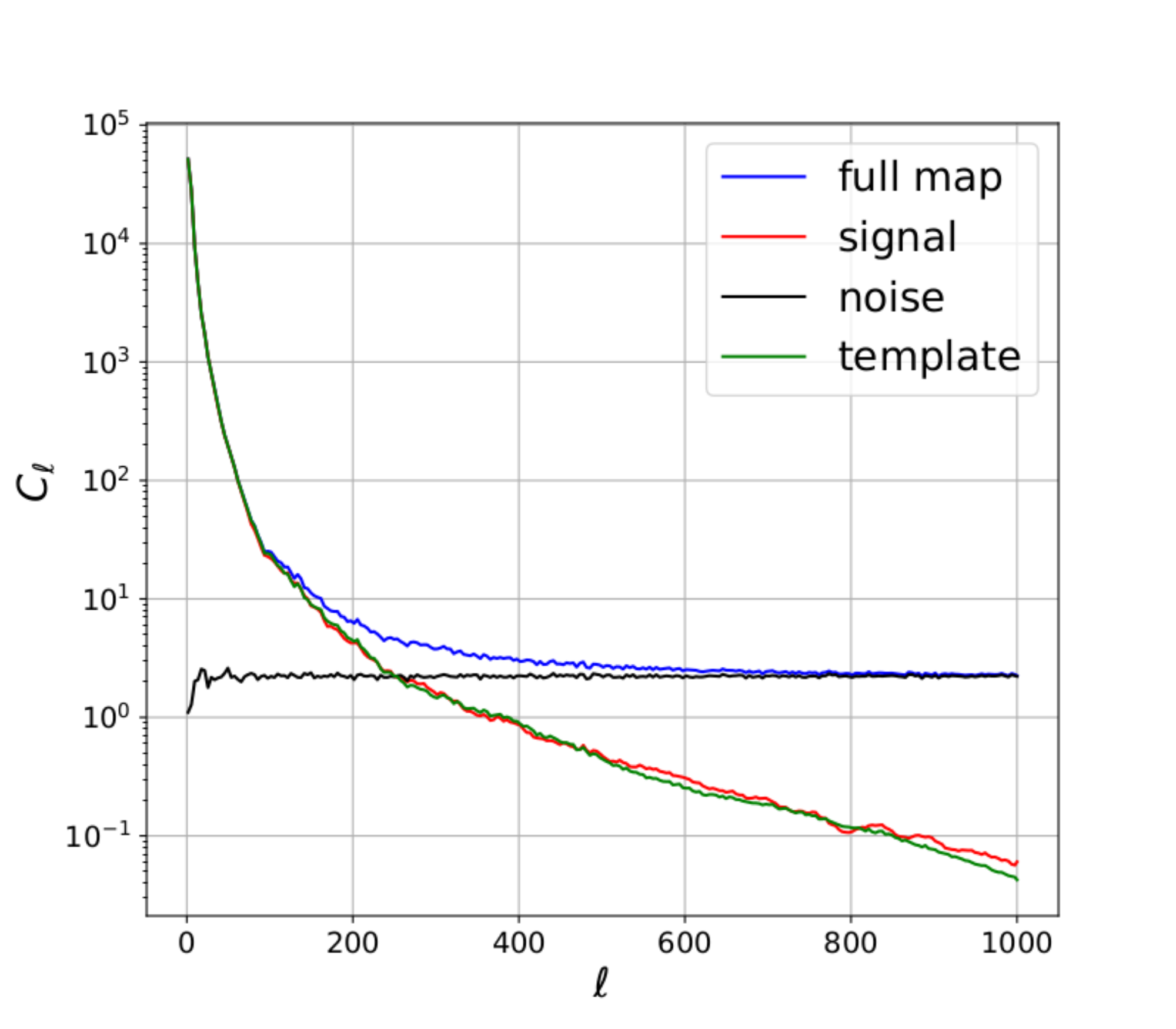}
    \includegraphics[width=0.45\textwidth]{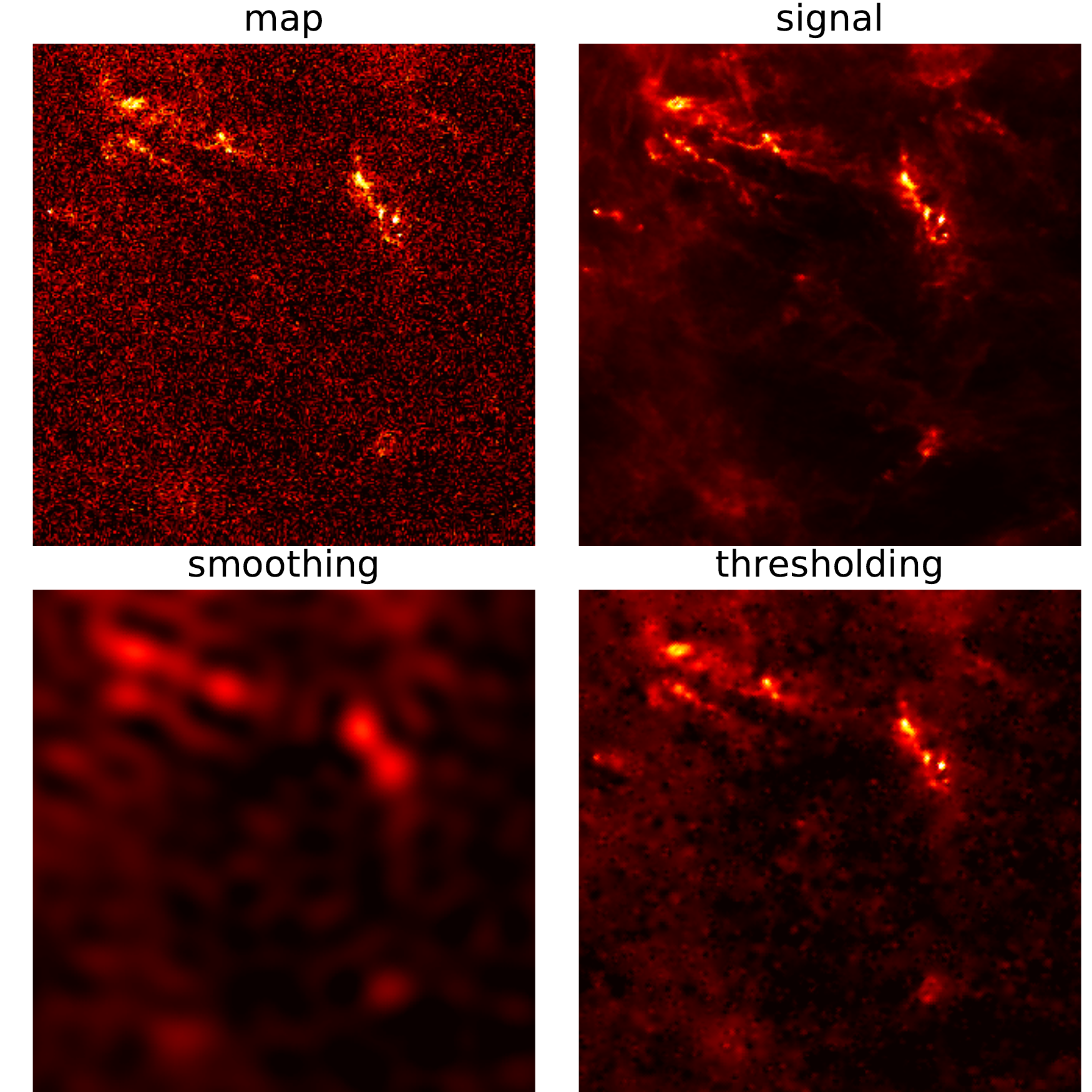}
    \caption{Left Panel: Power spectra of the input map (blue), of the noise (black), of the signal (red) and of the template recovered via thresholding. Right panel: a $30^{\circ} \times 30^{\circ}$ patch extracted from the input map. Clockwise from the top left: full input foreground signal $+$ noise, input foreground signal only, reconstructed foreground signal after thresholding, patch smoothed at signal dominated resolution.}
    \label{fig:temp_rec}
\end{figure}
\subsection{Template reconstruction}
\label{sec:temprec}
After this preliminary investigation, we now move on to show the actual performance of thresholding on the recovery of a signal template from noisy data.
In this analysis, we use as benchmark a map obtained from a foreground template at 200 GHz and an isotropic Gaussian noise realization, with HEALPix nside $512$. 
In order to highlight the denoising properties of thresholding, we set a very high noise level, so that a large portion of the available scales are noise dominated.
We run a soft thresholding algorithm on this synthetic map: the threshold is selected following the method introduced in \citep{doi:10.1080/01621459.1995.10476626}, based on the minimization of the Stein Unbiased Estimate of Risk (SURE), defined as, given a map $x$ and a thresholding operator $ST_\lambda(x)$:
\begin{equation}
   \operatorname{SURE}(x) = N\sigma^2 + ||ST_\lambda(x)-x||^2 + 2 \sigma^2 \sum_{i=1}^N \frac{\partial}{\partial x_i} (ST_\lambda(x)-x)_i,
\end{equation}
where $N$ is the dimension of the map $x$, $i$ runs over the pixels indices and, in the specific case of soft thresholding, the last summation corresponds simply to minus the number of thresholded coefficients.\\
As expected, thresholding is very effective in recovering the power of the input signal, deep into the noise dominated region.
This is shown in the left panel of figure \ref{fig:temp_rec}, where the angular power spectrum of the recovered template (in green) follows faithfully the one of the input signal (in red) at all scales and well below the noise level (in black). 
The excellent performance of thresholding is directly related to what discussed in the previous section. 
At high multipoles, corresponding to the noise dominated region in this example, the signal power is concentrated in a tiny number of very large $\beta_{jk}$, whereas the noise power is spread among all the coefficients.
In the right panel, we show the real space reconstruction of signal structures, compared to the noisy map and the input signal. 
We also show the map smoothed at the signal-dominated scale. The comparison with the thresholding results shows clearly how the latter removes a large part of the noise while also preserving the smallest structures.\\
In summary, this test shows that thresholding out small coefficients produces an almost complete suppression of the noise with only marginal loss of the signal power while maintaining the full resolution of the starting template.
Furthermore, as said before, the dimension of the data set is greatly reduced: the reconstruction presented here use only $7\%$ of the total number of coefficients.\par
\begin{figure}
    \centering
    \includegraphics[width=0.495\textwidth]{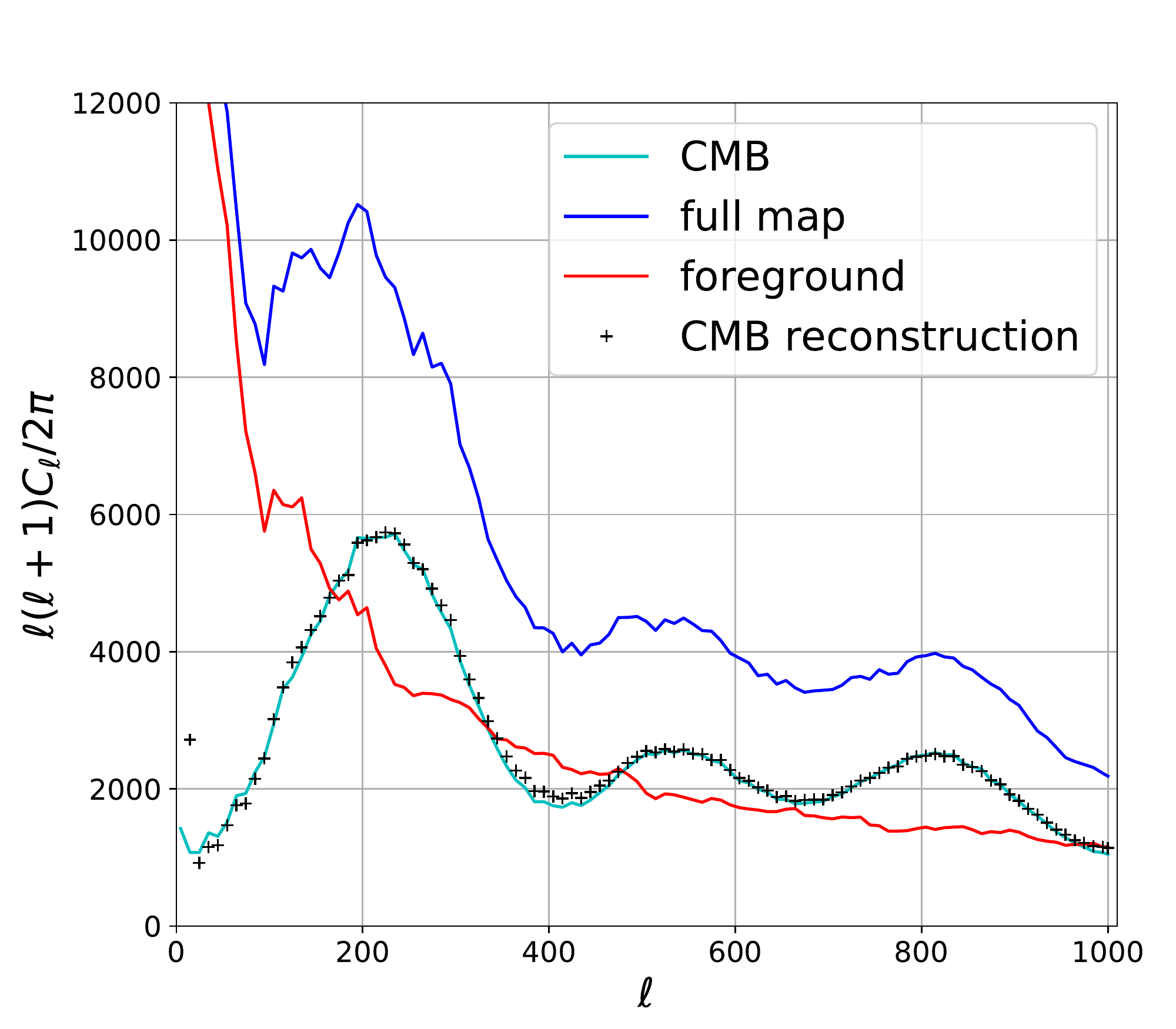}
    \includegraphics[width=0.495\textwidth]{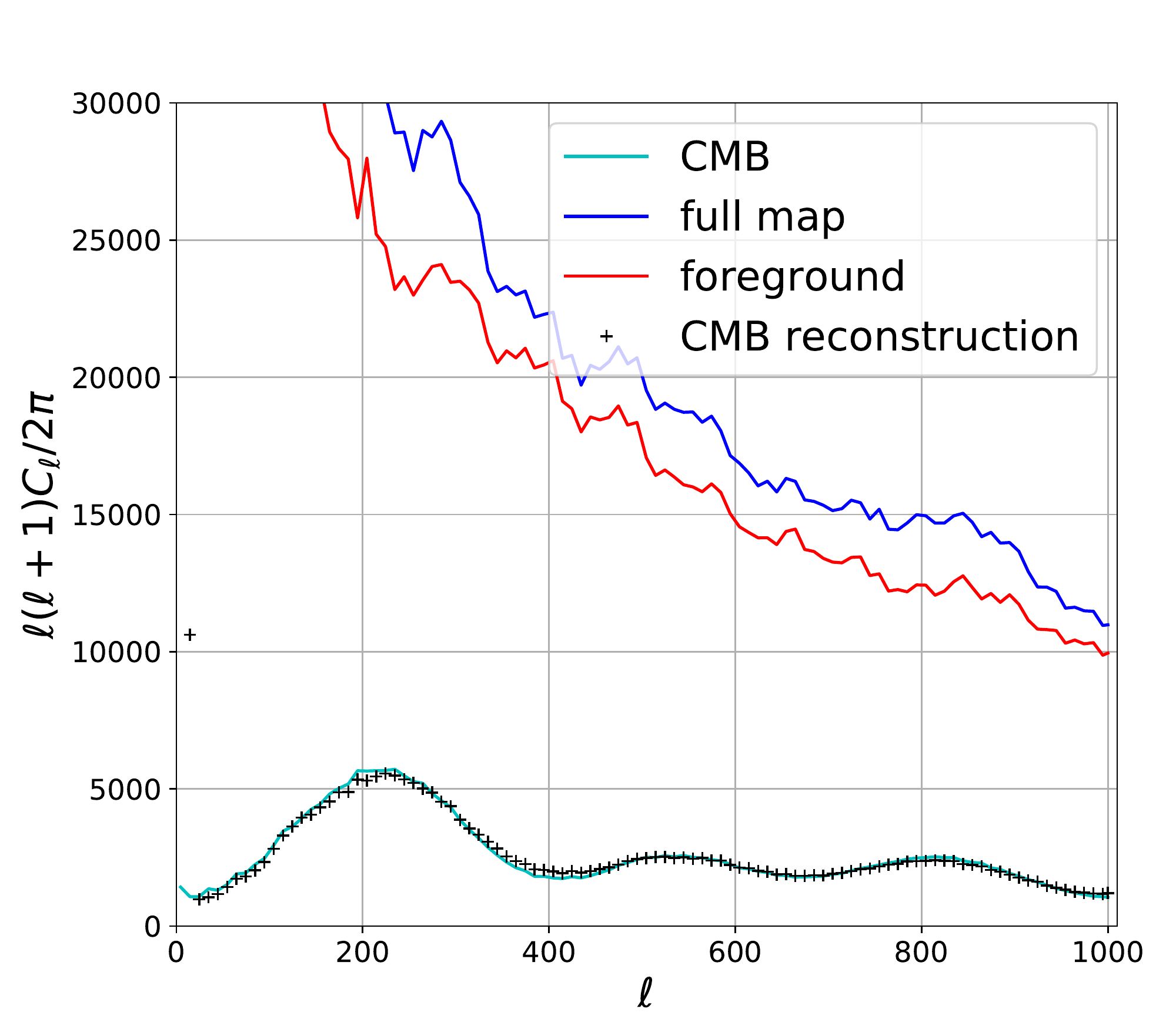}
\caption{Power spectra of the full map (blue) of the foreground (red), of the input CMB signal (cyan) and of the CMB reconstruction from the residuals of thresholding (black cross) for simulations at 143 GHz (left panel) and 217 GHz (right panel). }
\label{fig:cmbrec}
\end{figure}
Moreover, the denoising of a foreground template is not the only applications of these methods.
Needlet properties also allow us to separate the coherent (foreground) and stochastic (CMB, noise) components. Since we have just reconstructed a foreground template, using only $7\%$ of the map coefficients, it is natural to assume that the residual coefficients provide a reconstruction of the stochastic component, being it the noise (as in our previous example) or the CMB.
To check for this, we build a map from a foreground template and a CMB realization.
We are interested in the recovery of specific features of the CMB power spectrum, {\it e.g.} the acoustic peaks, thus, for the purpose of this test, we do not need to include noise in the map.
We show the results of this analysis in figure \ref{fig:cmbrec}, considering two realizations at 143 and 217 GHz respectively.
In this case, we used a hard thresholding algorithm implemented with a threshold selection criterion based on the minimization of the difference between the Gini index of the residuals with the expected value for a Gaussian field. 
In both cases we remove a small part of the galactic plane ($f_{sky}=90\%$).
Note that the size of the mask does not affect the results of thresholding.
We see that the power spectrum of the residuals (black cross) remarkably follows the CMB one (cyan line) for $\ell \gtrsim 20$ irrespective of the relative amplitude of the foreground component (red line).
On the other hand, at low multipoles the reconstruction completely fails. 
As we commented before, these multipoles represent the diffuse signal, that is not sparse and cannot be separated from the stochastic background with this technique.
\par
As a final interesting result we show that, in consequence of the sparsity of foregrounds, thresholding (and in particular soft thresholding) dominates all linear estimators in terms of mean square error, when we attempt to reconstruct a foreground template from a noisy realization.
The problem we have dealt with so far is that of estimating a foreground template $f$ from a single noisy observation $d=f+n$, where the noise is a Gaussian realization $n=N(0,{\bf I}\sigma)$, so that $d=N(f,{\bf I}\sigma)$. The usual least square solution would be simply $\hat f = d$. 
In contexts of this type, it however proven that the asymptotically optimal estimator is not least square but it is the so-called James-Stein (JS) estimator. More specifically, if we focus on linear estimators, and the dimension of the data-set is greater than 3, the JS estimator is known to provide be the lowest mean square error, always dominating the least square solution \citep{james1961,stein1956}. We will therefore focus here on this class of estimators to set our performance benchmark, which we will use to assess thresholding results later on.

Following the notation just defined, the JS estimator of the template f is defined as:
\begin{equation}
    \hat f^{JS}=\left( 1-\frac{(N-2)\sigma^2}{\sum_{i=1}^{N}d_i^2}\right)^+d
    \label{eq:JS}
\end{equation}
where $i$ runs over the mode indices in the chosen representation, N is the dimension of the data-set and the apex $+$ stand for the positive part. 
We develop a simple implementation of this estimator and test it in needlet, harmonic and real space. 
In real space, the application is straightforward: d is the map, $\sigma$ is the noise standard deviation, and $N$ is the number of pixels. 
For needlet (harmonic) space we apply the estimator adaptively, scale by scale and multipole by multipole, so that -- for each scale $j'$ (multipole $\ell$) -- $d$ in formula \ref{eq:JS} corresponds to  $\beta_{j'k}$ ($a_{\ell' m}$), while $\sigma^2$ corresponds to the needlet noise variance $\sigma_{j'}^2$ (noise power spectrum $C_{\ell'}$).
A general implementation would require the computation of the full noise covariance. However, in our idealized situation, where the noise is white and uncorrelated both in space and frequency, the solution we provide here is exact.
Note that, under the assumption that signal and noise are independent Gaussian random fields, the harmonic James-Stein estimator just described would be formally analogous to the Wiener filter solution.
Basically, in its scale/multipole dependent implementation, this estimator suppresses the noise dominated scales.\\
\begin{table}
\centering
\begin{tabular}{|c||c|c|c|c|c|c|}
\hline
{\bf method} &  & \multicolumn{3}{c}{James-Stein} & \multicolumn{2}{|c|}{Thresholding} \\
\cline{3-7}
 & smoothing & real space & needlets & harmonic & hard & soft  \\
\hline
{\bf rms} ($\sigma_{n}=27.10 {\rm mK}$) & $0.020\sigma_{n}$ & $0.026\sigma_{n}$ & $0.018\sigma_{n}$ & $0.023\sigma_{n}$ &  $0.020\sigma_{n}$ & $0.016\sigma_{n}$  \\ 
{\bf rms} ($\sigma_{n}=0.75 {\rm mK}$) & $0.295\sigma_{n}$ & $0.735\sigma_{n}$ & $0.259\sigma_{n}$ & $0.259\sigma_{n}$ &  $0.189\sigma_{n}$ & $0.170\sigma_{n}$  \\ 
{\bf rms} ($\sigma_{n}=0.28 {\rm mK}$) & $0.493\sigma_{n}$ & $0.925\sigma_{n}$ & $0.429\sigma_{n}$ & $0.426\sigma_{n}$ &  $0.315\sigma_{n}$ & $0.286\sigma_{n}$  \\ 
\hline
\end{tabular}
\caption{Root mean square errors of the real space template in unit of the noise standard deviation per pixel. The noise levels are set so that the power spectrum match the signal one at $\ell=20,250,600$ (top-down).}
\label{tab:rms}
\end{table}
As mentioned above, after applying the JS estimator to our case of interest, we compare its performance to thresholding.
In particular, we compare two thresholding algorithms: a soft thresholding where the threshold is selected minimizing SURE, and a hard thresholding based on the so-called universal threshold defined as:
\begin{equation}
    \lambda_j=\sigma_j\sqrt{2\log{N}} ,
\end{equation}
where $\sigma_j$ is the noise variance at the scale j and N in the number of needlet coefficients.
We find that, in the case of sparse signals, thresholding dominates ({\it i.e.} provides lower mean square error) even this estimator (and therefore any other linear estimator). 
We show this in Table \ref{tab:rms}, where we list the root mean square errors of the real space template with respect to the input signal, in units of $\sigma_{noise}$ for different estimators and noise levels for a 200GHz foreground template. 
The central row corresponds to the configuration used in the analysis shown in figure \ref{fig:temp_rec}.
As said earlier, we compare the results obtained with thresholding with the JS estimates in real, harmonic and needlet space, and with a simple  smoothing at signal dominated scales (respectively $\ell=20,250,600$ for the three configurations top-down).
Besides optimality issues, another advantage of needlet methods is that they are completely blind (under the assumption of white noise), whereas the other approaches assume knowledge of either the noise Power Spectrum or  $\sigma_{pixel}$. For the smoothing case, also some knowledge about the signal Power Spectrum is necessary, in order to set the scale of the low-pass filter.
On the contrary, in the needlet methods, the noise standard deviation (that appear both in SURE and in the JS shrinking factor) is computed directly from the data using the median absolute deviation (MAD) \footnote{The MAD times a given factor provides a robust estimator of the standard deviation in presence of outliers.} of the highest frequency layer, knowing that this is noise dominated.
As expected, thresholding always provides the lowest rms in these tests. In particular, soft thresholding with SURE based threshold selection achieves the best results in all cases. The hard thresholding estimator, based on the universal, threshold also provides good results, being slightly outperformed by the needlet JS estimator only in the noisiest case.
The improvement is always higher for higher level of noise for all, while the differences between them are more pronounced for lower noise.\\
Summarizing, our analyses so far have shown that thresholding achieves the best results in the extraction of foreground templates from noisy realizations. 
In the next section, we discuss some examples of applications of these techniques, in synergy with other component separation methods.
\begin{table}
\centering
\begin{tabular}{|l|c|c|c|}
\hline Channel & 140 & 220 & 240  \\ 
\hline Beam (arcmin) & 110 & 110 & 110 \\
\hline $\sigma$ ($\mu\mathrm{K}_{\rm CMB}$arcmin) & 30 & 40 & 80 \\
\hline
\end{tabular}
\caption{SWIPE specifications}
\label{tab:SWIPEspec}
\end{table}
\subsection{Synergy with other methods}

In this section, we present the results obtained by applying the techniques described in the previous sections to different simulated data-sets.
We generate two mock datasets using the PySM software \citep{2017MNRAS.469.2821T}, which mimic respectively observations of the {\it Planck} mission and of the SWIPE instrument of the forthcoming LSPE mission \citep{2012arXiv1208.0281T,2012SPIE.8452E..3FD,2016JLTP..184..527G}.\\
LSPE (Large Scale Polarization Explorer) is a forthcoming ASI mission aimed at the measurement of large scale CMB polarization fluctuations. 
It will scan a large region of the sky (20\%), with two instruments, SWIPE and STRIP. 
In this  work, we concentrate on the former, SWIPE (Short Wavelength Instrument for the Polarization Explorer).
SWIPE will measure CMB polarization in three frequency channels from a stratospheric balloon flying long-duration in the northern polar region during the winter night.\\
Due to its configurations, SWIPE represents a good benchmark for our methods: since it observes in just three different frequency channels, needlet thresholding can in principle significantly improve the results of blind component separation techniques as internal templates fitting and ILC. 
Planck simulations, instead, are useful to test the performance of the algorithms on the current state of the art CMB maps and, more in general, on full-sky data-sets with significant frequency coverage.
\begin{figure}
    \centering
    \includegraphics[width=0.495\textwidth]{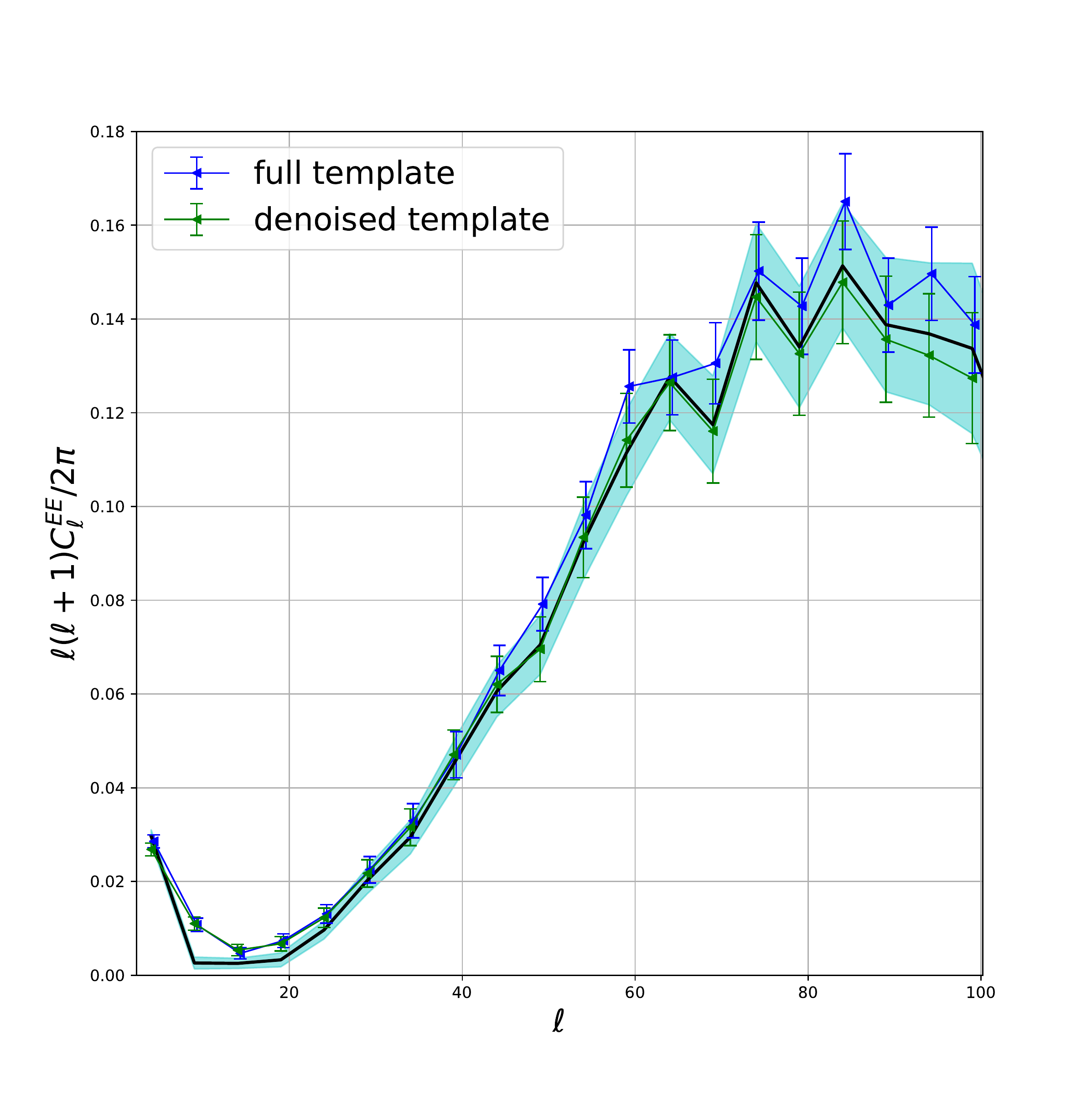}
    \includegraphics[width=0.495\textwidth]{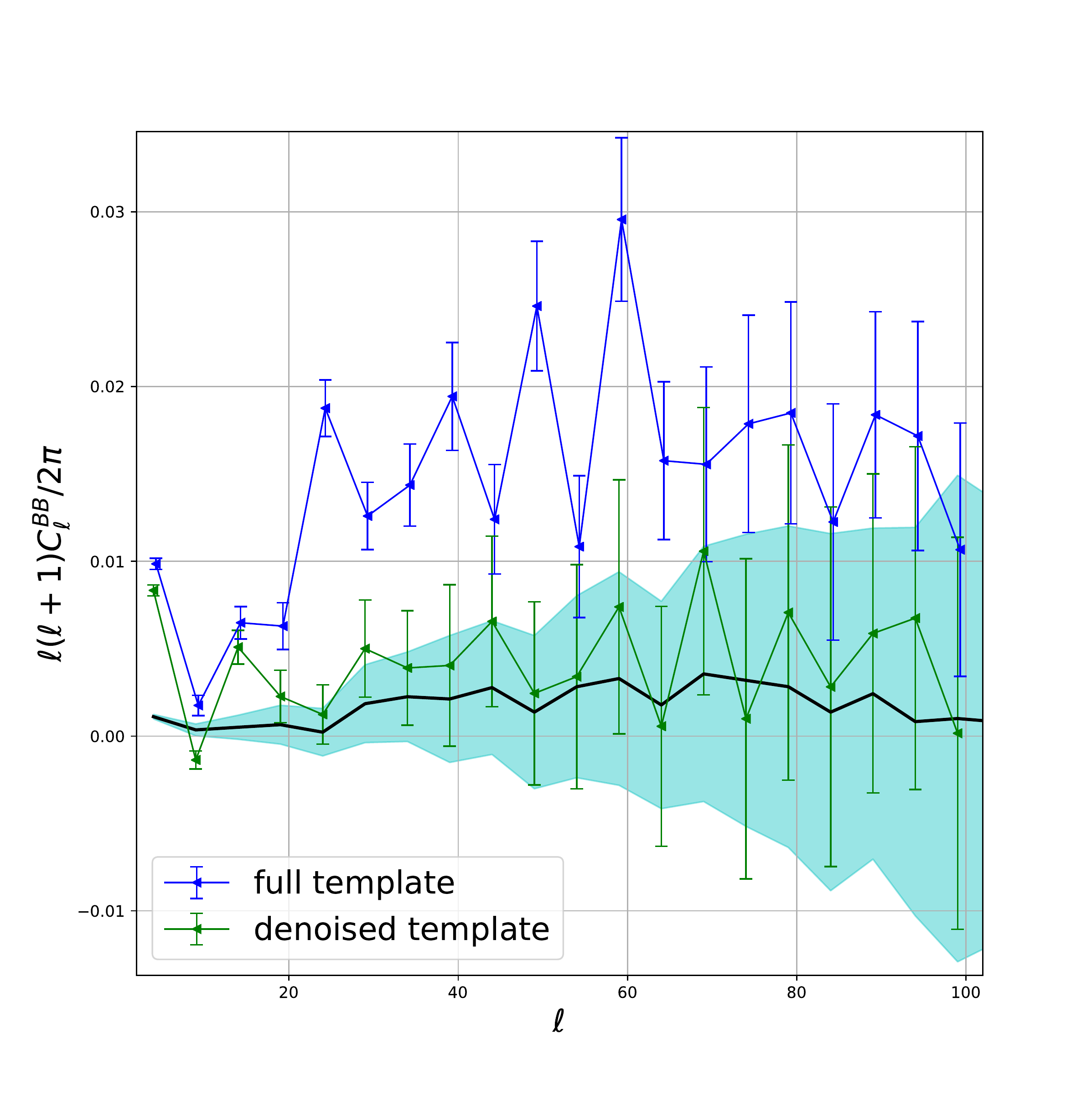}
    \caption{Power spectrum reconstruction with a template fitting algorithm in needlet space -- with and without thresholding -- for SWIPE simulations, corrected for the average noise contribution. The black line represents the average of the power spectra of the maps and the shaded blue area is its standard error. Green and blue lines represent the mean of the power spectra of the cleaned maps with and without thresholding respectively. The error bars are the respective standard errors.} \label{fig:tfneed}
\end{figure}
\subsubsection{Template Fitting}\label{sec:tfres}
In this section, we test the performance of our internal template fitting pipeline, equipped with the needlet thresholding method described before.
Our aim here is not that of presenting a full component separation pipeline, but rather that of quantifying the impact of the application of  thresholding techniques on simple template fitting procedures.\\
As a first test, we produce a set of 100 simulations of the three SWIPE channels, assuming the Planck cosmology and tensor-scalar ratio $r=0.1$, while the foregrounds are generated following the Planck sky model (see \citep{2017MNRAS.469.2821T} for reference).
Given the coarse angular resolution of the experiment under exam, simulations with HEALPix nside=128 are sufficient.
The noise is assumed to be white, Gaussian and isotropic. \\
With just three frequency bands, only one internal template can be used.
This regime provides therefore an ideal benchmark to verify the improvement from the pre-processing of the templates with our denoising algorithm.
The internal template is obtained from the difference between the 240 GHz and the 220 GHz channels, and it is used to clean the 140 GHz cosmological channel.
As a probe of the flexibility of this technique, we implement two different pipelines. 
The former strictly follows the steps described in section \ref{sec:method}. 
The latter still uses the same procedure to obtain the thresholded template, but the final fit is now performed in real space. \par
\begin{figure}
    \centering
    \includegraphics[width=0.495\textwidth]{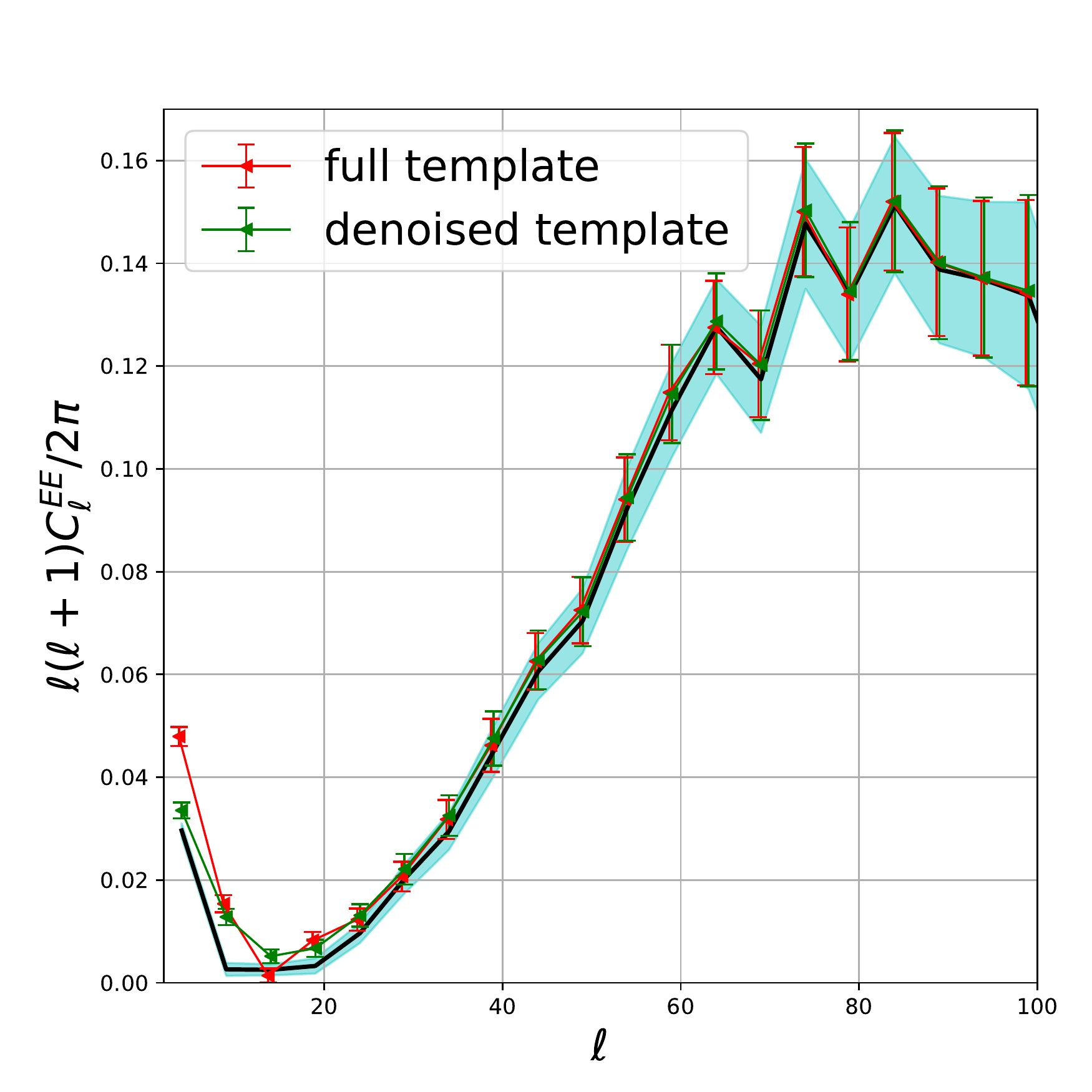}
    \includegraphics[width=0.495\textwidth]{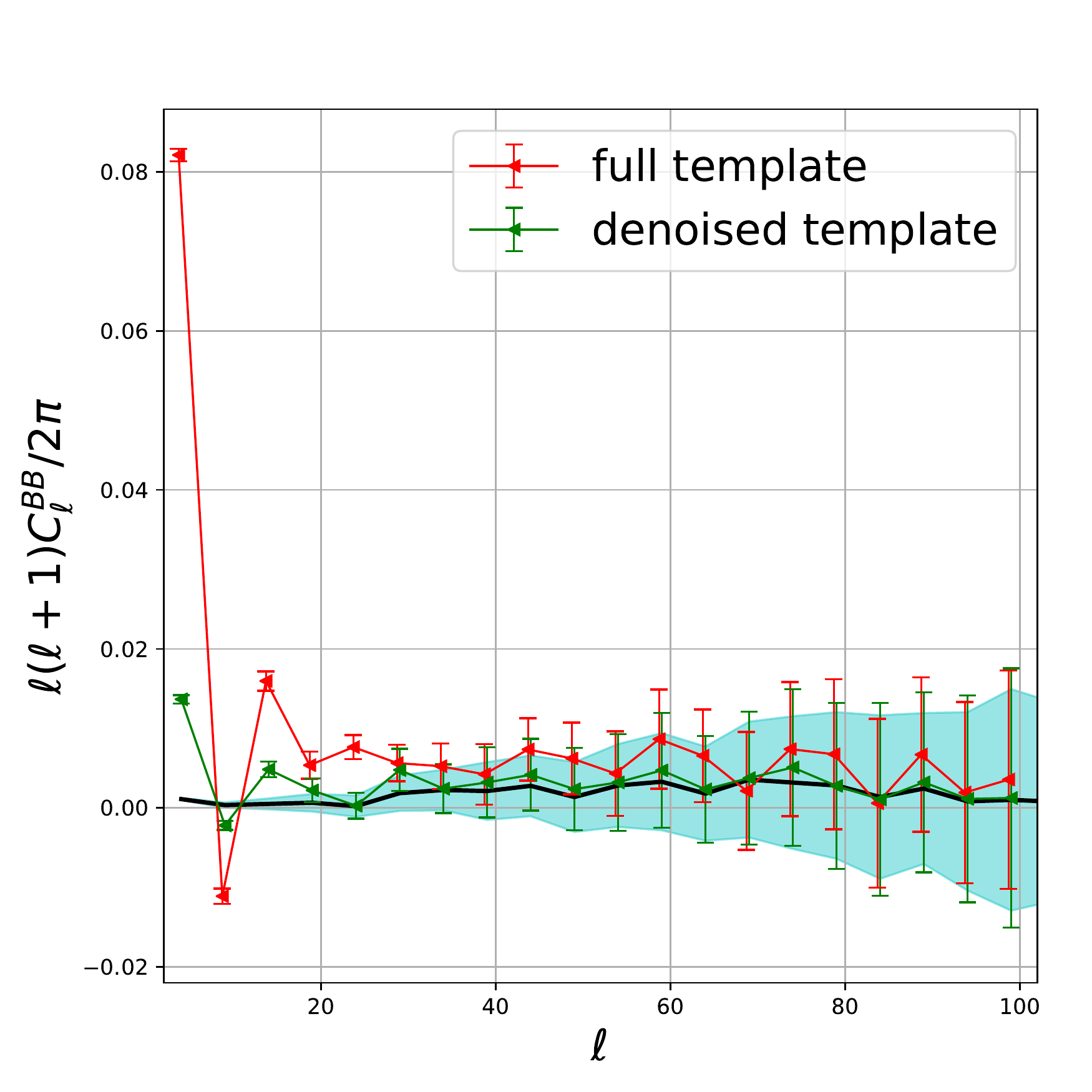}
    \caption{Power spectrum reconstruction using a template fitting algorithm in real space -- with and without thresholding -- for SWIPE simulations, corrected for the average noise contribution. The black line represents the average of the maps' power spectra and the shaded blue area is its standard error. Green and the red lines represent the mean of the power spectra of the cleaned maps with and without thresholding respectively; the error bars are the respective standard errors.}\label{fig:tfreal}
\end{figure}
\begin{figure}
    \centering
    \includegraphics[width=0.49\textwidth]{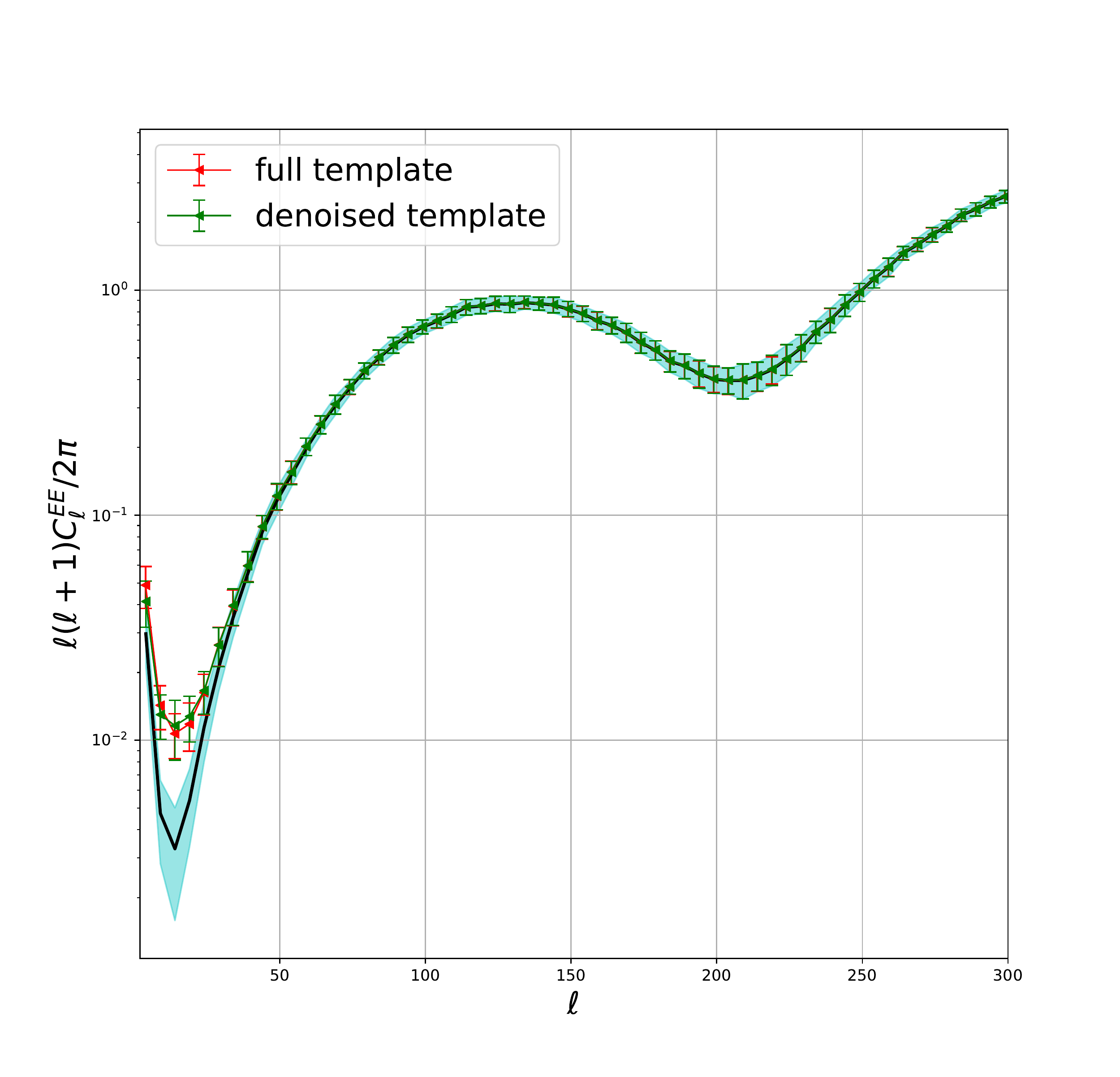}
    \includegraphics[width=0.49\textwidth]{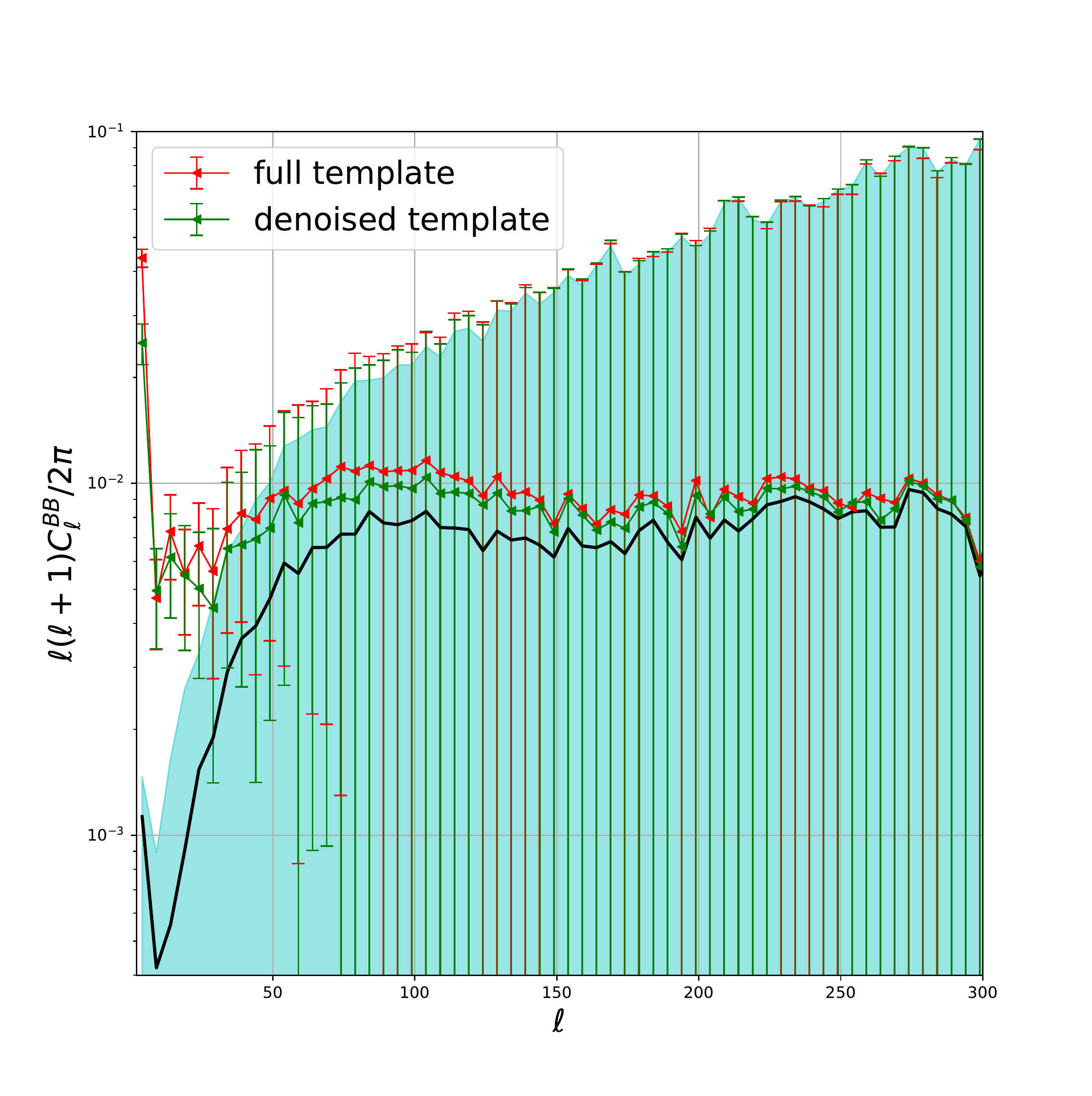}
    \caption{Same results as in figure \ref{fig:tfneed}, but for {\em Planck}-like simulations.} \label{fig:tfplanckreal}
\end{figure}

The overall approach to assess the performance of our method is quite straightforward: we decompose each set of polarization maps in E and B modes, followed by applying our template fitting algorithms with and without thresholding. 
We then measure EE and BB power spectra in all cases and use them as figure-of-merit, by comparing cleaned spectra with the input ones. \\
We show the results of this analysis in figure \ref{fig:tfneed} and \ref{fig:tfreal}.
We find that, for both techniques (needlet and real space fit), pre-cleaning the templates with the needlet thresholding provide noticeable improvements.
Especially in the case of B-mode reconstruction, where the noise dominate the templates, we found that preliminary denoising allows us to successfully recover the input power spectrum in a large portion of the multipole space under examination.
We repeat the same kind of analysis on simulations of Planck data. 
Planck, with 9 frequency channels, has a much larger frequency coverage than SWIPE, therefore we expect the effects of template thresholding to be less relevant in this case.
Given the preliminary level of this investigation, we do not produce full resolution simulations but we limit the maps to nside=256 and we restrict the analysis to lmax=500. Since we are looking at diffuse foregrounds on large scales, this does not alter significantly our final performance assessment.
For consistency, all the template are smoothed to match the Planck channel with lowest resolution, {i.e.} the 30 GHz channel with a 30 arcmin beam.
Our internal templates are built following the SEVEM pipeline described in \citep{2018arXiv180706208P}.
 In figure \ref{fig:tfplanckreal} we show results for the cleaning of the 147 GHz channel. 
Following the Planck team, we use as internal templates the difference between the channels: (30-44)GHz, (217-100)GHz, (353-217)GHz.
The first traces the synchrotron, while the last two will trace the dust emission. \\
Our results are obtained on a set of 50 simulation; as expected in this case, the impact of thresholding is much less relevant than in the LSPE case. 
However, some improvement can still be noticed in the B-mode spectrum reconstruction, where the signal to noise is very low.
Let us note here that, in principle, other denoising techniques  can be applied to this problem, with a good performance in terms of the mean square error. For example, if we focus on the B-mode maps, a simple smoothing of the template at very low resolution can still produce an accurate fit. This is because - using the foreground model of our simulations (based on PySM)- the result is mostly driven by the first few multipoles, where the foreground amplitude is well above the template noise.
However, even in cases where the error improvement by using thresholding is only marginal, we stress the fact that soft thresholding does maintain a number of important advantages.
The main advantage is that thresholding preserves the original resolution of the maps after denoising, thus allowing us to reconstruct and study the actual structure of the template on smaller scales in real observations. All this can be achieved at the same time with a huge amount of data compression, as discussed and tested in detail in the previous section. This last property is of particular interest in the context of template fitting, because a smaller data-set requires to invert a smaller covariance matrix, with a remarkable computational gain. 
No other technique among those we analyzed allows us to achieve this combined result, i.e.,  optimality, no loss of resolution and data compression. 
If we focus for example on the denoising method described in our previous section, we already commented how smoothing the maps has the obvious drawback of removing potentially relevant information at small scales. The JS estimator, on the other hand, preserve resolution, but does not allow for data compression.  
\subsubsection{ILC}\label{sec:ILCres}
\begin{figure}
    \centering
    \includegraphics[width=0.49\textwidth]{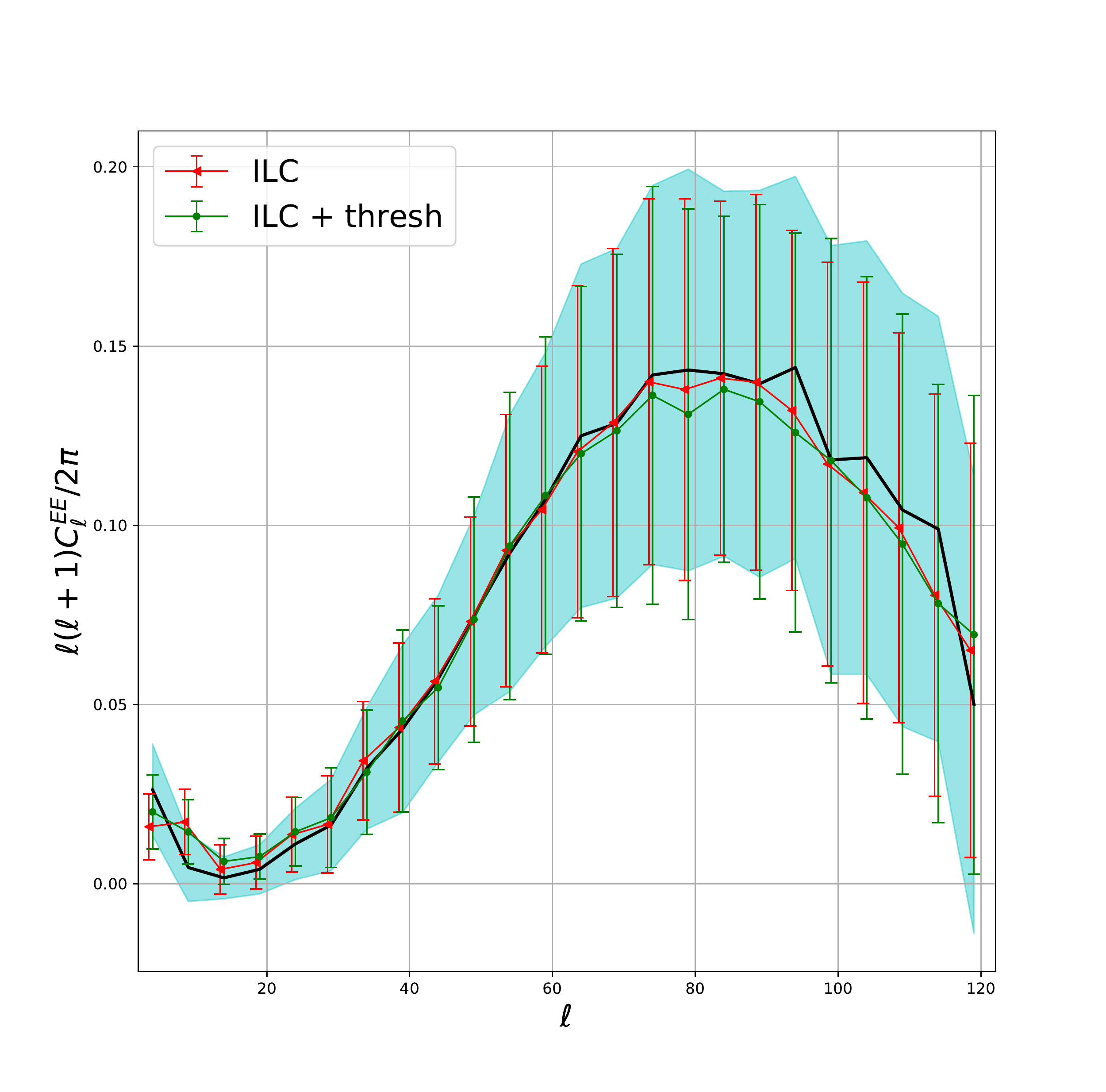}
    \includegraphics[width=0.49\textwidth]{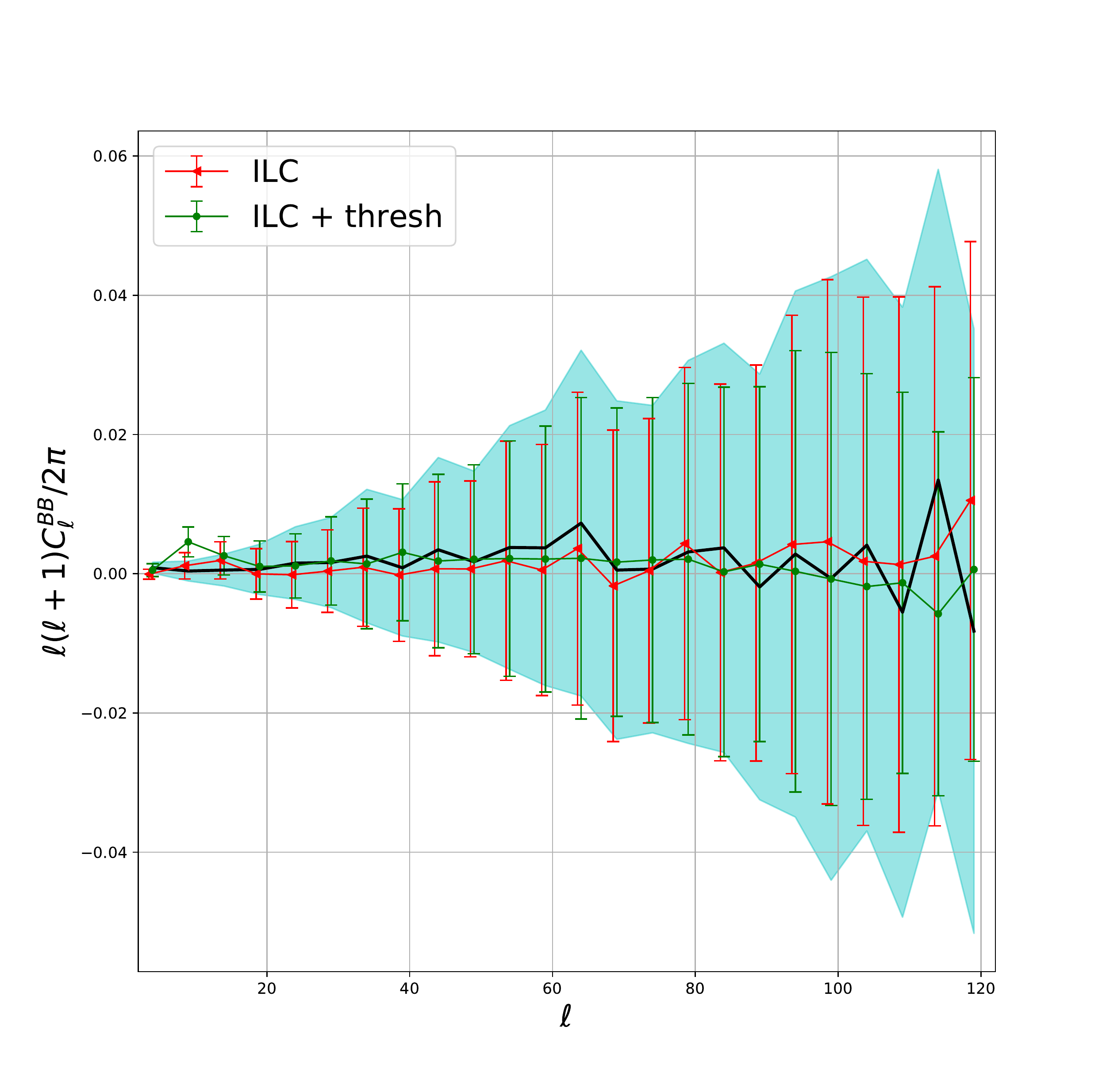}
    \caption{Power spectrum reconstruction with ILC with and without thresholding for SWIPE simulations, corrected for the average noise contribution. 
    The conventions are the same as in previous figures: the average power spectrum from input simulations is represented in black, with the shaded blue area showing its standard error. Green and red lines show the average power spectra of the cleaned map with and without thresholding respectively.}\label{fig:ilclspe}
\end{figure}

\begin{figure}
    \centering
    \includegraphics[width=0.49\textwidth]{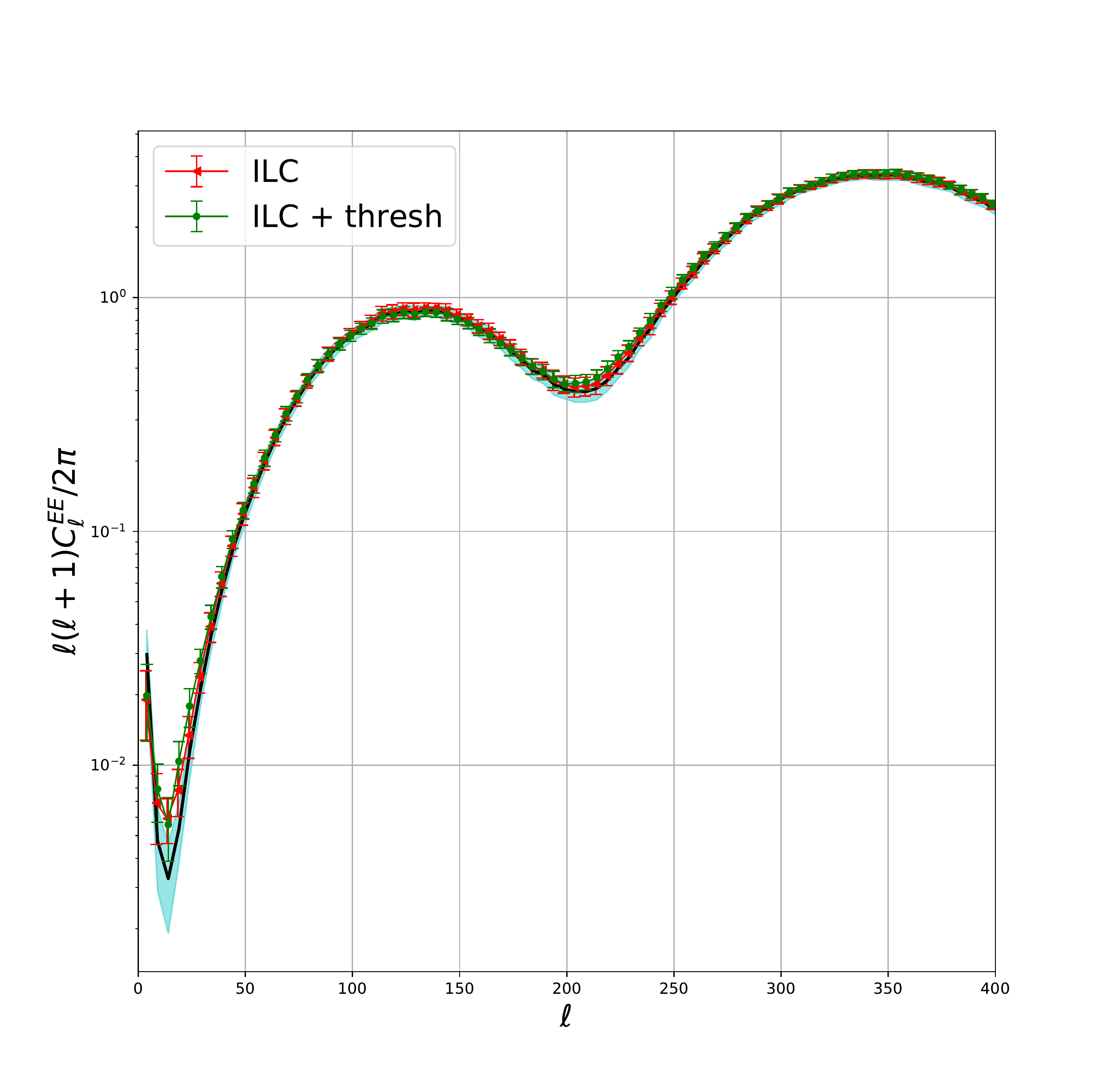}

    \caption{Same results as shown in figure \ref{fig:ilclspe}, but for EE Power Spectrum of {\em Planck}-like simulations}\label{fig:ilcplanck}
\end{figure}

Here we present a similar analysis as in the previous section but we focus on ILC techniques rather than template fitting.
We work on the same set of simulations described before. 
We consider a needlet space ILC approach, implemented via the algorithm described in section \ref{sec:method}, where the map needlet coefficients are thresholded before combining the channels.
We compare this technique with a needlet space ILC where the weights are left free to vary between scales.
Before applying this ``standard'' needlet ILC with no thresholding, we mask the galactic plane with the Planck component separation common mask in polarization.
On the other hand, our method does not need any masking since the thresholding automatically remove the most contaminated regions.
We want to verify how our blind threshold selection criterion based on the minimization of anisotropy performs with respect to a standard, ``a priori'' masking procedure.\\
Figures \ref{fig:ilclspe} and \ref{fig:ilcplanck} show the results for LSPE and Planck respectively.
We find that the results are comparable, proving that our technique performs well in selecting the foreground contaminated regions.
We point out here that our currently implemented ILC technique can be improved in several ways, {\it e.g.} changing the weight in different areas of the sky.
However, we did not introduce these additional refinements since -- as for the previous template-fitting analysis -- they are not required for what we are strictly interested about in this work, namely the specific impact of thresholding on the cleaning procedure.\\
These results show how this technique can act as an automatic, blind masking method, with similar performance as the standard real space mask. 
It is relevant to notice that this approach allows recovering the input power spectrum without external assumptions on the contaminated areas of the sky (i.e., the masked part of the sky is recovered internally via thresholding, and no externally generated galactic mask is required). 
We also stress again that the role of thresholding, in this case, is conceptually completely different from the denoising performed in the template fitting algorithm. This is an additional probe of the large flexibility of this method.
\section{Conclusions}
\label{sec:conc} 
In this paper, we showed several applications of needlet thresholding techniques to the problem of CMB component separation.\par
The unifying idea behind our study is that of exploiting sparsity of foreground components in the needlet representation, as a tool to separate foregrounds from the stochastic background by exploiting their peculiar morphological features (anisotropy, non-Gaussianity).\par
In the first part of our analysis, we tested explicitly on simulations that our needlet expansion of foreground templates is sparse, by using Gini coefficients as a measure of sparsity. 
We then showed how thresholding allows reconstructing a noisy template with high accuracy, up to small scales, well below the noise level. 
We also made a comparison between different denoising techniques, showing that for our purposes, needlet thresholding has the best performance in terms of reconstruction accuracy, while preserving the full resolution of the templates and at the same time achieving strong data compression.\par
After this investigation, in the second part of our study we implemented specific needlet thresholding procedures 
as extensions of existing component separation techniques. We then verified whether and in which situations this could improve the final CMB reconstruction. 
To this purpose, we focused on two well-known component separation procedures, namely ILC and template-fitting, considering simulated data sets and using as figure of merit the reconstruction of the input CMB polarization power spectrum. We compared bot soft- and hard-thresholding schemes and developed different procedures to set the optimal threshold level.\par
In the case of ILC, the role of thresholding is that of "pre-cleaning" single channels, before combining all the frequencies. This captures information on the foreground spatial distribution, which complements spectral frequency information and can in principle lead to a more accurate foreground cleaning procedure. In the case of template fitting, we have instead already discussed how thresholding is a powerful denoising method for internal templates.\par
After applying our algorithms to realistic simulations of different experimental setups, we found in practice that thresholding can be useful in experiments with few frequency channels, in conditions of low signal-to-noise. This is logical, since in these cases the original internal foreground templates are very noisy and the small frequency coverage reduces the accuracy of the standard approaches.\par
The best performance of thresholding in our tests are in particular found when considering a template fitting technique in an LSPE like experiment, especially for B-mode. Our ILC-thresholded algorithm, where we set the threshold level to maximize isotropy, gives instead similar results to standard ILC. Similarly, as anticipated, only marginal improvements are obtained in both cases for a Planck-like experiment with many frequency bands.\par
After the preliminary exploration discussed in this paper, we will therefore focus in a future work on developing in detail a full thresholding-based, needlet template-fitting pipeline.
We will also explore the performance of this approach in a different context, namely for foreground cleaning and template reconstruction in intensity mapping experiments.

\acknowledgments
FO's and ML's work is supported by the University of Padova under the STARS Grants programme CoGITO, Cosmology beyond Gaussianity, Inference, Theory and Observations, funding: 150 k\euro{} . 
DM acknowledges the MIUR Excellence Department Project awarded to the Department of Mathematics, University of Rome Tor Vergata, CUP E83C18000100006.
DB acknowledges partial financial support by ASI Grant No. 2016-24-H.0.
FO, ML, DB, CB and DP acknowledge support from the ASI-COSMOS Network (cosmosnet.it) and from the INDARK INFN Initiative (web.infn.it/CSN4/IS/Linea5/InDark)
\bibliographystyle{JHEP.bst}
\bibliography{main.bib}

\end{document}